\documentclass[journal]{IEEEtran}
\usepackage{bbm}
\usepackage{array}
\usepackage{dcolumn}
\usepackage{epsfig}
\usepackage[intlimits]{amsmath}
\usepackage{amsmath, amsfonts, yhmath, bm}
\usepackage{amssymb}
\usepackage{psfrag}
\usepackage{color,soul}
\usepackage[dvipsnames]{xcolor}
\usepackage[normalem]{ulem}
\usepackage{enumerate}
\usepackage{stackengine}
\usepackage[noadjust]{cite}
\usepackage{graphicx}
\usepackage{multirow}
\usepackage[caption=false,font=footnotesize]{subfig}
\usepackage{etoolbox}
\usepackage{tcolorbox}
\usepackage{float}
\usepackage{dsfont}
\usepackage{tikz}
\usetikzlibrary{arrows}
\usetikzlibrary{calc}
\usetikzlibrary{decorations.pathmorphing}
\usetikzlibrary{arrows.meta,positioning,fit}
\usepackage{comment}
\usepackage{ifthen}
\let\CMmathcal\mathcal      %
\usepackage{eucal}         %
\let\mathcal\CMmathcal     %

\usepackage{hyperref}
\hypersetup{
    colorlinks,
    linkcolor={red!50!black},
    citecolor={blue!50!black},
    urlcolor={blue!80!black}
}

\usepackage{balance}
\usepackage{url}
\usepackage[inline]{enumitem} %

\usetikzlibrary{backgrounds, positioning, fit}
\usepackage{pgfplots}
\pgfplotsset{compat=newest}  %
\usepgfplotslibrary{colormaps}

\usepackage{mathtools} %

\usepackage{vmr-symbols-vecbold}
\usepackage{standard-macros}
\PassOptionsToPackage{hyphens}{url}

\newcommand{\defeq}{\mathrel{\mkern-0.25mu=}}
\DeclareSymbolFontAlphabet{\amsmathbb}{AMSb}%

\newcommand{\lro}[1]{\lefto({#1}\right)}																%

\newcommand{\lr}[1]{\left({#1}\right)}																%
\newcommand{\lrb}[1]{\left \lbrace {#1} \right \rbrace}															%

\safemath{\dopplerspread}{B_D}																								%
\safemath{\delayspread}{T_D}																									%
\safemath{\nc}{n\sub{c}}																										%
\safemath{\nf}{n\sub{f}}																										%
\safemath{\efa}{p\sub{sc}}
\safemath{\efb}{p\sub{cs}}
\safemath{\ef}{\epsilon\sub{f}	}
\safemath{\nd}{n\sub{d}}																										%
\safemath{\ntx}{n\sub{t}} 																											%
\safemath{\nrx}{n\sub{r}}																											%
\safemath{\ntxt}{\tilde{n\sub{t}}}																											%
\safemath{\cb}{\ensuremath{L}} 																								%
\safemath{\cl}{\ensuremath{n}} 																								%
\safemath{\txanto}{{\ensuremath{\tilde{m}_t}}} 																		%
\safemath{\cs}{M} 																														%
\safemath{\idPustm}{\ensuremath{S_{k}}}
\safemath{\error}{\ensuremath{\epsilon}} 																				%
\safemath{\eexp}{\ensuremath{\mathcal{E}}} 																			%
\safemath{\nsubc}{n\sub{s}}			 																						%
\safemath{\nofdm}{n\sub{o}} 																									%
\safemath{\bc}{\ensuremath{B_c}} 																							%
\safemath{\ts}{\ensuremath{T_s}} 																							%
\safemath{\nrb}{\ensuremath{n_{rb}}} 																						%
\safemath{\rul}{\ensuremath{\rho\sub{ul}}}
\safemath{\rdl}{\ensuremath{\rho\sub{dl}}}

\safemath{\nres}{\ell}
\safemath{\nr}{n\sub{r}}
\safemath{\maxk}{M^*\lr{\nres, \nsubc, \nofdm, \epsilon, \rho}}
\safemath{\Rmax}{R^*}%
\safemath{\Emin}{E\sub{b}^*/N_0}%
\safemath{\Eminf}{\frac{E\sub{b}^*}{N_0}}
\safemath{\np}{\ensuremath{n\sub{p}}}
\safemath{\ndf}{\ensuremath{\bar{n}\sub{d}}}
\safemath{\npf}{\ensuremath{\bar{n}\sub{p}}}
\safemath{\code}{\ensuremath{\mathcal{C}}}
\safemath{\err}{\ensuremath{\epsilon}}
\safemath{\rp}{\ensuremath{\rho\sub{p}}}
\safemath{\rd}{\ensuremath{\rho\sub{d}}}
\safemath{\cohtime}{\ensuremath{T\sub{c}}}
\safemath{\cohbw}{\ensuremath{B\sub{c}}}
\safemath{\nmax}{\ensuremath{\ell\sub{m}}}
\safemath{\ntot}{\ensuremath{n\sub{tot}}}
\safemath{\nul}{\ensuremath{n\sub{ul}}}
\safemath{\ndl}{\ensuremath{n\sub{dl}}}

\safemath{\yp}{\ensuremath{\randvecy_{\nu}^{(\text{p})}}}
\safemath{\yd}{\ensuremath{\randvecy_{\nu}^{(\text{d})}}}
\safemath{\ypd}{\ensuremath{\vecy_{\nu}^{(\text{p})}}}
\safemath{\ydd}{\ensuremath{\vecy_{\nu}^{(\text{d})}}}

\safemath{\ypf}{\ensuremath{\bar{\randvecy}_{\nu}^{(\text{p})}}}
\safemath{\ydf}{\ensuremath{\bar{\randvecy}_{\nu}^{(\text{d})}}}
\safemath{\ypdf}{\ensuremath{\bar{\vecy}_{\nu}^{(\text{p})}}}
\safemath{\yddf}{\ensuremath{\bar{\vecy}_{\nu}^{(\text{d})}}}

\safemath{\xp}{\ensuremath{\vecx^{(\text{p})}}}
\safemath{\xd}{\ensuremath{\randvecx_{\nu}^{(\text{d})}}}
\safemath{\xdd}{\ensuremath{\vecx_{\nu}^{(\text{d})}}}

\safemath{\xpf}{\ensuremath{\bar{\vecx}^{(\text{p})}}}
\safemath{\xdf}{\ensuremath{\bar{\randvecx}_{\nu}^{(\text{d})}}}
\safemath{\xddf}{\ensuremath{\bar{\vecx}_{\nu}^{(\text{d})}}}

\safemath{\xdb}{\ensuremath{\overline{\randvecx}^{(\text{d})}}}
\safemath{\Pxd}{\ensuremath{P_{\randvecx^{(\text{d})}}}}

\safemath{\xpbar}{\ensuremath{\overline{\matX}^{(\text{p})}}}
\safemath{\xdbar}{\ensuremath{\overline{\randmatX}^{(\text{d})}}}

\safemath{\xdv}{\ensuremath{\randvecx^{(\text{d})}}}
\safemath{\xdbarv}{\ensuremath{\overline{\randvecx}^{(\text{d})}}}
\safemath{\ydv}{\ensuremath{\randvecy^{(\text{d})}}}

\safemath{\xdr}{\ensuremath{\matX^{(\text{d})}}}

\safemath{\ttx}{\ensuremath{\tau\sub{tx}}}
\safemath{\trx}{\ensuremath{\tau\sub{rx}}}
\safemath{\ack}{\ensuremath{\mathrm{s}}}
\safemath{\nack}{\ensuremath{\mathrm{c}}}

\safemath{\mI}{\ensuremath{i\lro{\randvecy ; \randvecx}}} 				%

\safemath{\randveca}{\bm{A}}
\safemath{\randvecb}{\bm{B}}
\safemath{\randvecc}{\bm{C}}
\safemath{\randvecd}{\bm{D}}
\safemath{\randvece}{\bm{E}}
\safemath{\randvecf}{\bm{F}}
\safemath{\randvecg}{\bm{G}}
\safemath{\randvech}{\bm{H}}
\safemath{\randveci}{\bm{I}}
\safemath{\randvecj}{\bm{J}}
\safemath{\randveck}{\bm{K}}
\safemath{\randvecl}{\bm{L}}
\safemath{\randvecm}{\bm{M}}
\safemath{\randvecn}{\bm{N}}
\safemath{\randveco}{\bm{O}}
\safemath{\randvecp}{\bm{P}}
\safemath{\randvecq}{\bm{Q}}
\safemath{\randvecr}{\bm{R}}
\safemath{\randvecs}{\bm{S}}
\safemath{\randvect}{\bm{T}}
\safemath{\randvecu}{\bm{U}}
\safemath{\randvecv}{\bm{V}}
\safemath{\randvecw}{\bm{W}}
\safemath{\randvecx}{\bm{X}}
\safemath{\randvecy}{\bm{Y}}
\safemath{\randvecz}{\bm{Z}}
\safemath{\randvecphi}{\bm{\Phi}}

\safemath{\randmatA}{\amsmathbb{A}}
\safemath{\randmatB}{\amsmathbb{B}}
\safemath{\randmatC}{\amsmathbb{C}}
\safemath{\randmatD}{\amsmathbb{D}}
\safemath{\randmatE}{\amsmathbb{E}}
\safemath{\randmatF}{\amsmathbb{F}}
\safemath{\randmatG}{\amsmathbb{G}}
\safemath{\randmatH}{\amsmathbb{H}}
\safemath{\randmatI}{\amsmathbb{I}}
\safemath{\randmatJ}{\amsmathbb{J}}
\safemath{\randmatK}{\amsmathbb{K}}
\safemath{\randmatL}{\amsmathbb{L}}
\safemath{\randmatM}{\amsmathbb{M}}
\safemath{\randmatN}{\amsmathbb{N}}
\safemath{\randmatO}{\amsmathbb{O}}
\safemath{\randmatP}{\amsmathbb{P}}
\safemath{\randmatQ}{\amsmathbb{Q}}
\safemath{\randmatR}{\amsmathbb{R}}
\safemath{\randmatS}{\amsmathbb{S}}
\safemath{\randmatT}{\amsmathbb{T}}
\safemath{\randmatU}{\amsmathbb{U}}
\safemath{\randmatV}{\amsmathbb{V}}
\safemath{\randmatW}{\amsmathbb{W}}
\safemath{\randmatX}{\amsmathbb{X}}
\safemath{\randmatY}{\amsmathbb{Y}}
\safemath{\randmatZ}{\amsmathbb{Z}}
\safemath{\randmatSigma}{\mathbb{\Sigma}}
\safemath{\randmatPhi}{\mathbb{\Phi}}
\safemath{\randmatLambda}{\mathbb{\Lambda}}

\safemath{\matSigma}{\bm{\Sigma}}
\safemath{\matPhi}{\bm{\Phi}}
\safemath{\matLambda}{\bm{\Lambda}}

\newcommand{\crlb}{\text{\tiny CRLB}}
\newcommand{\icrlb}{\text{\tiny iCRLB}}

\newcommand {\isr} {\mathtt{ISR}}

\newcommand{\ratedist}{\text{\tiny RD}}

\safemath{\blkdiag}{\operatorname{blkdiag}}
\safemath{\vecop}{\operatorname{vec}}

\definecolor{mitgrayback}{cmyk}{0.06, 0.05, 0.05, 0.0}
\definecolor{lightblack}{cmyk}{0.3, 0.3, 0.3, 0.25}
\definecolor{lightgray}{cmyk}{0.02, 0, 0, 0.0}
\definecolor{uppboundcolor}{RGB}{0,102,204}

\newcommand{\delra}[1]{}

\makeatletter
\newsavebox\myboxA
\newsavebox\myboxB
\newlength\mylenA

\newcommand*\mybar[2][0.75]{%
	\sbox{\myboxA}{$\m@th#2$}%
	\setbox\myboxB\null%
	\ht\myboxB=\ht\myboxA%
	\dp\myboxB=\dp\myboxA%
	\wd\myboxB=#1\wd\myboxA%
	\sbox\myboxB{$\m@th\overline{\copy\myboxB}$}%
	\setlength\mylenA{\the\wd\myboxA}%
	\addtolength\mylenA{-\the\wd\myboxB}%
	\ifdim\wd\myboxB<\wd\myboxA%
	\rlap{\hskip 0.5\mylenA\usebox\myboxB}{\usebox\myboxA}%
	\else
	\hskip -0.5\mylenA\rlap{\usebox\myboxA}{\hskip 0.5\mylenA\usebox\myboxB}%
	\fi}
\makeatother

\makeatletter
\renewcommand*{\defeq}{\mathrel{\rlap{%
			\raisebox{0.3ex}{$\m@th\cdot$}}%
		\raisebox{-0.3ex}{$\m@th\cdot$}}%
	=}
\makeatother

\newtheorem{remark}{Remark}

\newtheorem{proposition}{Proposition}

\newtheorem{definition}{Definition}
\newtheorem{example}{Example}
\begin{document}

\title{Distributed Blind Source Separation from Compressed Mixtures}

\author{Amir~Weiss
\thanks{A. Weiss is with The Alexander Kofkin Faculty of Engineering, Bar-Ilan University, Ramat Gan, 5290002, Israel, e-mail: amir.weiss@biu.ac.il.}}%

\maketitle

\begin{abstract}
Blind source separation (BSS) is increasingly needed in sensing architectures where mixtures are acquired by multiple remote devices but the inference itself is carried out centrally. In such systems, the sensors often cannot transmit their raw observations since the communication links are rate-limited, energy-limited, or both. The fusion center therefore must recover the latent sources from compressed versions of the mixtures. This setting raises questions that classical BSS models do not address directly: whether source identifiability survives compression, {how separation accuracy changes under a specified compression model}, and how the separation procedure should be modified to account for the statistical model induced by the compression stage. In this work, we develop a framework for distributed BSS from compressed mixtures that addresses these issues at three levels. First, we formalize identifiability under compression and establish a data-processing viewpoint showing that compression cannot introduce distinguishability that is absent before compression. Second, for mixtures of Gaussian sources with general covariance structures, we derive {an idealized Gaussian rate--distortion-based Cram\'er--Rao lower bound benchmark for the interference-to-source ratio}. Third, we develop practical compression-aware separation schemes based on subtractively dithered quantization together with debiasing at the fusion center for both fourth-order-cumulant-based and second-order-statistics-based methods. Simulation experiments demonstrate the resulting tradeoffs between communication rate and separation performance and the gains obtained by compression-aware separation in representative distributed sensing scenarios.
\end{abstract}


\IEEEpeerreviewmaketitle

\vspace{-0.3cm}
\section{Introduction}\label{sec:introduction}

Blind source separation (BSS) deals with the recovery of latent signals from their observed mixtures, with minimal prior knowledge on the mixing mechanism or the sources. As one of the central problems in statistical signal processing, it concerns various applications spanning audio and speech enhancement, biomedical signal analysis, array processing, and neuroimaging (e.g., \cite{Vincent2018Audio,mckeown1998analysis,Calhoun2001fMRIICA}). A particularly influential BSS paradigm is independent component analysis (ICA), which leverages statistical independence among sources to identify a demixing transform and separate the components \cite{Comon1994ICAconcept,Cardoso1998BSSprinciples,HyvarinenOja2000ICA,BellSejnowski1995Infomax}.

Classical ICA methodology often relies on non-Gaussianity of the latent sources and the implied higher-order statistics (HOS) to resolve the mixing matrix, leading to practical algorithms based on information maximization and cumulants, including Infomax and joint approximate diagonalization of eigen-matrices (JADE) \cite{BellSejnowski1995Infomax,cardoso1993blind}. When sources exhibit a nondegenerate temporal structure, second-order statistics (SOS) become informative: distinct autocorrelation or spectral signatures provide ``temporal diversity'' that enables separation even for Gaussian sources, motivating SOS-based methods such as the second-order blind identification (SOBI) algorithm \cite{Belouchrani1997SOBI}.

Complementing these algorithmic developments, likelihood-based formulations---e.g., maximum likelihood (ML) and quasi-ML (QML)---provide statistically grounded baselines and performance characterizations in temporally diverse models \cite{yeredor2010blind,yeredor2012sedjoco,weiss2018performance,weiss2018consistency}. Importantly, these developments predominantly assume that a processor has access to the mixtures in (effectively) full precision, so that the relevant statistics (e.g., HOS-based cumulants, covariances and likelihood terms) can be estimated directly from the observed waveforms.

In modern sensing deployments, however, mixture acquisition is frequently distributed across spatially separated sensors, while inference is performed at a fusion center (FC). Communication, energy, and latency constraints in wireless sensor networks and Internet-of-Things platforms often make raw waveform streaming infeasible. Consequently, the FC typically receives only compressed (possibly quantized) descriptions of the local mixtures, and must perform BSS from lossy, distributed messages rather than from the uncompressed data.

Previous work on distributed and decentralized ICA is primarily motivated by scalability, privacy, or multi-site collaboration, but the communication model in these works is typically iteration-based and effectively real-valued, involving the exchange of gradients, moments, or model updates, rather than explicit per-sensor finite-rate descriptions of the observed mixtures (e.g.,~\cite{Baker2019DecentralizedICA,ImtiazSarwate2019DPDecentralizedICA,musluoglu2025distributed}). Finite-rate representations of continuous-valued observations are, of course, classical in quantization theory and source coding \cite{GrayNeuhoff1998,Schuchman1964Dither}. Moreover, for Gaussian processes, rate--distortion (RD) theory provides a principled reconstruction-oriented benchmark for characterizing the tradeoff between compression rate and signal fidelity \cite{CoverThomas2006}. However, the BSS literature largely lacks a unified treatment that connects \emph{distributed finite-rate compression of mixture waveforms} to (i) \emph{identifiability} of the mixing model at the FC, (ii) \emph{accuracy {benchmarks}} for estimating the mixing (or demixing) parameters and the separated signals, and (iii) \emph{practical separation methods} that are matched to the statistical model induced by compression. This gap is particularly pronounced in SOS-driven Gaussian ICA, where separation hinges on preserving temporal diversity that may be distorted or destroyed by naive quantization.

{A central consequence of the compressed-observation formulation is that parameter-independent sensor encoders cannot create statistical distinguishability that is absent from the uncompressed mixtures. In other words, compression can only preserve or destroy the equivalence classes required for BSS, and it cannot ``repair" a nonidentifiable source model. This data-processing viewpoint motivates both the destructive sign-quantization example and the structure-preserving dithered construction developed in Section~\ref{subsec:gausssourcesos}.}

{Communication-constrained BSS is also complementary to recent distributed sensing and integrated sensing and communication formulations that optimize beamforming, antenna positions, synchronization, or robustness to imperfect channel state information~\cite{xiu2026cooperative,xiu2026metarl,xiu2026robust,lyu2025movable,xiu2025lowaltitude}. Those formulations modify the physical sensing and communication architecture. Here, the instantaneous mixing geometry is fixed, and the question is instead which source-separation information survives sensor-side compression and how the FC should account for the induced statistical model. In particular, spatial reconfiguration and compression-aware BSS act on different design layers: a reconfigurable array may improve the conditioning or observability of the physical mixture, whereas compression-aware BSS seeks to retain and exploit source diversity after the observations have been encoded. The present model excludes synchronization errors, uncertain channels, and jointly optimized sensing links. Integrating these effects with rate allocation and separation is left for future work.}

This paper addresses distributed BSS under explicit per-sensor rate constraints by combining a statistical identifiability viewpoint with information-theoretic and estimation-theoretic benchmarks, and with practical compression--separation procedures. Our main contributions are as follows:
\begin{itemize}

\item We formulate \emph{distributed BSS from compressed mixtures} via per-sensor encoders and a FC-side decoder, and define the notion of \emph{equidentifiability}, i.e., identifiability up to an equivalence relation that captures the permutation and scaling ambiguities of ICA. To expose how compression interacts with SOS-based identifiability, we present canonical examples of equidentifiability-destructive and equidentifiability-preserving compression rules, leveraging subtractively dithered quantization as a tractable additive-noise model at the FC;

\item We {use} a data-processing principle showing that \emph{compression cannot create equidentifiability} when it is absent in the uncompressed model. {This observation anchors the compressed-BSS equivalence relation and focuses the substantive design question on which encoder invariances destroy, and which compressed statistics preserve, the original ICA equivalence classes.} We specialize the discussion to Gaussian ICA where SOS govern separability;

\item We adopt a RD-optimal Gaussian test-channel model to parameterize the compressed observations as a function of per-sensor rates, derive a Cram\'er--Rao lower bound (CRLB) for estimating the mixing parameters, and translate this bound into a benchmark on the separation quality metric interference-to-source ratio (ISR); and

\item We propose practical fixed-rate compression--separation schemes based on subtractively dithered uniform quantization and explicit debiasing of the induced statistics at the FC, yielding (i) a fourth-order-cumulant-based pipeline for temporally independent and identically distributed (i.i.d.)\ sources (e.g., JADE~\cite{cardoso1993blind}), and (ii) a SOS-based pipeline for wide-sense stationary (WSS) sources (e.g., SOBI~\cite{Belouchrani1997SOBI} or QML~\cite{weiss2018consistency}).

\end{itemize}

The rest of the paper is organized as follows. Section~\ref{sec:problemformulation} presents the signal and compression models and formalizes the performance criterion. Section~\ref{sec:identifiability} studies identifiability under compression and develops a data-processing viewpoint. Section~\ref{sec:benchmarks} introduces oracle-normalized benchmarks and notions of separation consistency under lossy compression. Section~\ref{sec:rd_model} develops the RD-based Gaussian compression model and derives {a Gaussian RD-based CRLB benchmark}, including a CRLB-induced bound on the ISR. Section~\ref{sec:practical_methods} presents practical compression--separation algorithms. Section~\ref{sec:sim_results} reports simulation results corroborating our theoretical results, and Section~\ref{sec:conclusion} concludes the paper with broader research directions and open challenges.

\vspace{-0.2cm}
\subsection{Notations}\label{subsec:notations}
Scalars are denoted by lowercase letters, vectors by bold lowercase letters, and matrices by bold uppercase letters. Random scalars and vectors are written in sans-serif font, e.g., $\rndx$ and $\rvecx$, whereas deterministic matrices are written in bold font, e.g., $\matA$. Unless stated otherwise, all quantities are real-valued. For $M\in\naturals$, we write $\setS_M\triangleq\{1,\ldots,M\}$, and use $\reals$, $\naturals$, and $\integers$ for the sets of real, natural, and integer numbers, respectively. A bracketed index, e.g., $\rndx[n]$ or $\rvecx[n]$, denotes the sample at discrete time~$n$, and a.s.\ denotes almost surely. We write $\rveca \perp \rvecb$ to say that $\rveca,\rvecb$ are statistically independent.

Expectation, variance, covariance, and probability are denoted by $\Exop[\cdot]$, $\Varop(\cdot)$, $\Covop(\cdot)$, and $\Prob(\cdot)$, respectively, and $\normal(\mu,\sigma^2)$ denotes the Gaussian distribution with mean $\mu$ and variance $\sigma^2$. The transpose, inverse and Kronecker product are denoted by $\tp{(\cdot)}$, $\inv{(\cdot)}$ and $\otimes$, respectively. We write $\diag(\cdot)$, $\blkdiag\{\cdot\}$, $\vecop(\cdot)$, and $\Tr(\cdot)$ for the (vector to) diagonal matrix, (matrices to) block-diagonal concatenation, (matrix) vectorization, and trace operators, respectively.

\vspace{-0.2cm}
\section{Problem Formulation}\label{sec:problemformulation}

Let $\rvecs[n] \triangleq \tp{[\rnds_1[n]\,\cdots\,\rnds_K[n]]} \in \reals^{K\times 1}$ denote the vector of $K\in\naturals$ latent sources of interest, where $\{\rnds_k[n]\}_{k\in\setS_K}$ are mutually independent zero-mean random processes. The temporal SOS of the sources are captured by their autocorrelation sequences,
\begin{equation}
    C_{\rnds_k}[n,m] \triangleq \Exop\!\big[\rnds_k[n]\rnds_k[m]\big],
    \quad
    k\in\setS_K,
    \quad
    n,m\in\integers,
    \label{eq:source_cov_general}
\end{equation}
or, for finite $N$-length sequences, by their covariance matrices,
\begin{equation}\label{eq:covsourcedef}
    \matC_{\rvecs_k}\triangleq \Covop(\rvecs_k)\in\reals^{N\times N},
\end{equation}
where, by definition, $C_{\rnds_k}[n,m]$ is the $(n,m)$-th element of \eqref{eq:covsourcedef}. Note that we do not impose stationarity at this stage.

We assume that there are $L$ spatially separated sensors, and for simplicity, we focus on the square mixing matrix case as a baseline, in which the number of sensors is equal to the number of sources, i.e., $L=K$. The $\ell$-th sensor observes a noisy linear (instantaneous) mixture of all $K$ sources,
\begin{equation}
    \rndx_\ell[n] = \sum_{k=1}^{K} a_{\ell k}\rnds_k[n] + \rndw_\ell[n],
    \quad
    \ell\in\setS_K,
    \quad
    n\in\integers,
    \label{eq:sensor_model}
\end{equation}
where $a_{\ell k}\in\reals$ are unknown deterministic mixing coefficients, and $\{\rndw_\ell[n]\}_{\ell\in\setS_K}$ are mutually independent zero-mean Gaussian noise processes. We assume that the noises are both spatially and temporally uncorrelated, and statistically independent of the sources, which implies
\begin{align}
    \forall\, n,m\in\integers: \; \begin{cases}
    \Exop\!\left[\rndw_\ell[n]\rndw_\ell[m]\right]&=\sigma_{\rndw_{\ell}}^2\cdot\mathbbm{1}_{n=m}, \quad \forall\ell\in\setS_K,
    \\
    \Exop\!\left[\rndw_\ell[n]\rndw_r[m]\right]&=0, \quad \forall\ell\neq r\in\setS_K,
    \\
    \Exop\!\left[\rndw_\ell[n]\rnds_k[m]\right]&=0,
    \quad
    \forall\, \ell,k\in\setS_K.
    \end{cases}
    \label{eq:noise_uncorr}
\end{align}
Let $\matA\in\reals^{K\times K}$ denote the unknown mixing matrix whose $(\ell,k)$-th entry is $a_{\ell k}$, and define the mixture and noise vectors $\rvecx[n] \triangleq \tp{[\rndx_1[n]\,\cdots\,\rndx_K[n]]} \in \reals^{K\times 1}$ and $\rvecw[n] \triangleq \tp{[\rndw_1[n]\,\cdots\,\rndw_K[n]]} \in \reals^{K\times 1}$, respectively, so that the network-wide measurement model reads
\begin{equation}
    \rvecx[n] = \matA\rvecs[n] + \rvecw[n],
    \quad
    n\in\integers.
    \label{eq:vector_model}
\end{equation}

Now, fix a blocklength $N\in\naturals$ and define, for each $\ell\in\setS_K$ and $k\in\setS_K$,
\begin{align}
    \rvecs_k &\triangleq \tp{[\rnds_k[1]\,\cdots\,\rnds_k[N]]} \in \reals^{N\times 1},\\
    \rvecw_\ell &\triangleq \tp{[\rndw_\ell[1]\,\cdots\,\rndw_\ell[N]]} \in \reals^{N\times 1},\\
    \rvecx_\ell &\triangleq \tp{[\rndx_\ell[1]\,\cdots\,\rndx_\ell[N]]} \in \reals^{N\times 1}.\label{eq:block_vectors_def}
\end{align}
If we stack across sensors and sources to have
\begin{align}
    \rvecs &\triangleq \tp{\left[\tp{\rvecs_1}\,\cdots\,\tp{\rvecs_K}\right]} \in \reals^{KN\times 1},\\
    \rvecw &\triangleq \tp{\left[\tp{\rvecw_1}\,\cdots\,\tp{\rvecw_K}\right]} \in \reals^{KN\times 1},\\
    \rvecx &\triangleq \tp{\left[\tp{\rvecx_1}\,\cdots\,\tp{\rvecx_K}\right]} \in \reals^{KN\times 1},\label{eq:stacked_blocks_def}
\end{align}
then~\eqref{eq:vector_model} can be equivalently expressed as the block model
\begin{equation}
    \rvecx = (\matA\otimes\matI_N)\rvecs + \rvecw\in\reals^{KN\times 1}.
    \label{eq:block_model}
\end{equation}

Let $\matC_{\rvecs}\triangleq \Covop(\rvecs)\in\reals^{KN\times KN}$ and $\matC_{\rvecw}\triangleq \Covop(\rvecw)\in\reals^{KN\times KN}$ denote the finite-block covariance matrices. By independence of the sources, $\matC_{\rvecs}$ is block-diagonal across the $K$ source blocks,
\begin{equation}
    \matC_{\rvecs} = \blkdiag\{\matC_{\rvecs_1},\ldots,\matC_{\rvecs_K}\},
    \label{eq:Cs_blockdiag}
\end{equation}
and under~\eqref{eq:noise_uncorr}, $\matC_{\rvecw}$ is block-diagonal across sensors,
\begin{align}
    \matC_{\rvecw} &= \blkdiag\{\matC_{\rvecw_1},\ldots,\matC_{\rvecw_K}\},
    \label{eq:Cw_blockdiag}
\end{align}
where $\matC_{\rvecw_\ell} \triangleq \Covop(\rvecw_\ell)\in\reals^{N\times N}$. Consequently,
\begin{align}
    \matC_{\rvecx} &\triangleq \Covop(\rvecx)\\
    &=
    (\matA\otimes\matI_N)\matC_{\rvecs}(\tp{\matA}\otimes\matI_N) + \matC_{\rvecw}\in\reals^{KN\times KN}.
    \label{eq:Cx_finite_block}
\end{align}
When the sources and noise are additionally WSS, $\matC_{\rvecs_k}$ and $\matC_{\rvecw_\ell}$ become Toeplitz. This specialization will be used later to obtain interpretable frequency-domain characterizations.

{We note that our treatment is restricted to instantaneous linear mixtures. Convolutive mixtures, while important and relevant for many applications, would require handling frequency-dependent mixing, filtering ambiguities, and rate allocation across both sensors and frequency bins. Thus, in order to focus on first principles, we leave their treatment for future work.}

\vspace{-0.3cm}
\subsection{Distributed Compression Model}

Each sensor communicates with the FC over a rate-limited but otherwise noiseless digital link. Sensor $\ell\in\setS_K$ maps its local observation block $\rvecx_\ell\in\reals^{N\times 1}$ into a message
\begin{equation}
    \rndm_\ell = f_\ell(\rvecx_\ell),
    \quad
    \rndm_\ell \in \setM_\ell,
    \label{eq:encoder_general}
\end{equation}
where $\setM_\ell$ is the codebook of the $\ell$-th encoder $f_\ell:\reals^{N\times 1}\to\setM_{\ell}$, and the latter may be deterministic or randomized. The rate constraint of the encoders is expressed as
\begin{equation}
    \log_2|\setM_\ell| \le NR_\ell,
    \quad
    R_\ell\in\positivereals,
    \quad
    \ell\in\setS_K,
    \label{eq:rate_constraint}
\end{equation}
where $R_{\ell}$ is the maximal rate of the $\ell$-th sensor. We focus on the setting where encoders operate independently across sensors with no access to mixtures observed by other sensors and no access to shared randomness unless stated otherwise.
{The subtractively dithered constructions presented below in Section~\ref{sec:practical_methods} are the stated exception: sensor $\ell$ and the FC share a parameter-independent pseudorandom seed that generates the dither sequence. All probability laws and data-processing statements may equivalently be conditioned on this seed.}

The FC receives the messages $\{\rndm_\ell\}_{\ell\in\setS_K}$ and needs to separate the sources. When it is convenient to represent the received information in signal form, we write
\begin{equation}
    \widehat{\rvecx}_\ell \triangleq h_{\ell}(\rndm_\ell) \in \reals^{N\times 1},
    \quad
    \ell\in\setS_K,
    \label{eq:reconstruction_general}
\end{equation}
for a reconstruction mapping (decoder) $h_{\ell}:\setM_{\ell}\to\reals^{N\times 1}$, and further define
\begin{equation}
    \widehat{\rvecx} \triangleq \tp{\left[\tp{\widehat{\rvecx}_1}\,\cdots\,\tp{\widehat{\rvecx}_K}\right]} \in \reals^{KN\times 1}.
    \label{eq:xhat_stack_def}
\end{equation}
No other structural assumptions are imposed on the encoders $\{f_\ell\}_{\ell\in\setS_K}$ or decoders $\{h_{\ell}\}_{\ell\in\setS_K}$ beyond locality and~\eqref{eq:rate_constraint}.

\vspace{-0.3cm}
\subsection{Separation Performance Criterion}

While the FC generally aims to separate the sources, there are different, suitably selected objectives in different settings. Thus, in the fully-blind case, i.e., when the sources' distributions are unknown, and in particular their covariances $\{\matC_{\rvecs_k}\}$ are unknown, we quantify separation quality via the ISR. Consider any demixing matrix $\widehat{\rmatB}\in\reals^{K\times K}$, an estimate of the true demixing matrix $\matB\triangleq\inv{\matA}$, formed at the FC based on $\{\rndm_\ell\}_{\ell\in\setS_K}$, and define the gain (``demixing-mixing") matrix,
\begin{equation}
    \rmatG \triangleq \widehat{\rmatB}\matA \in \reals^{K\times K}.
    \label{eq:gain_matrix_def}
\end{equation}
The component-wise ISR is defined as (see, e.g., \cite[Eq.(2)]{yeredor2010blind})
\begin{equation}
    \isr_{k\ell}(\rmatG) \triangleq \Exop\!\left[\rndg_{k\ell}^2\right]\cdot
    \frac{\Exop\!\left[\tp{\rvecs_\ell}\rvecs_\ell\right]}{\Exop\!\left[\tp{\rvecs_k}\rvecs_k\right]}\in\positivereals,
    \;\;
    \forall k\neq \ell \in\setS_K,
    \label{eq:isr_k_def}
\end{equation}
where $\rndg_{k\ell}$ is the $(k,\ell)$-th entry of $\rmatG$, and \eqref{eq:isr_k_def} represents the relative residual energy of the $\ell$-th source in the reconstruction of the $k$-th source. We further define the average ISR as
\begin{equation}
    \isr(\rmatG) \triangleq \frac{1}{K(K-1)}\sum_{\substack{k,\ell=1 \\ k\neq \ell}}^{K}\isr_{k\ell}.
    \label{eq:isr_def}
\end{equation}

{We use ISR since it isolates the source-leakage component of BSS, is invariant to the standard permutation and scaling ambiguities, and is linked directly to the demixing-matrix CRLB derived in the sequel. For an end-to-end audio application, scale-invariant signal-to-distortion ratio (e.g.~\cite{le2019sdr}) or perceptual measures would be useful complementary metrics. However, they also reflect observation noise, waveform-reconstruction distortion, and the chosen output scaling or filtering convention. These effects are distinct from the source leakage targeted by the present mixing-matrix analysis.}

Using the ISR as our separation quality criterion, in the following sections we work towards three main objectives:
\begin{itemize}
    \item Establish conditions for \emph{separability}, namely, identifiability of $\matA$ (up to permutation and scaling) from compressed distributed measurements of the mixtures;

    \item Derive a {Gaussian RD-based CRLB} {benchmark} on separation performance (in terms of ISR); and

    \item Develop blind separation algorithms that operate on the compressed mixtures, which are ``good" in the sense of minimizing \eqref{eq:isr_k_def} and \eqref{eq:isr_def}, and w.r.t.\ the derived bound under appropriate regimes.
\end{itemize}

{For convenience, Table~\ref{tab:principal_notation} summarizes the
principal notation used throughout the paper.}
{\begin{table*}[t]
    \centering
    \caption{{Principal observation, compression, and separation notation.}}
    \label{tab:principal_notation}
    \footnotesize
    \setlength{\tabcolsep}{5pt}
    \renewcommand{\arraystretch}{1.08}
    \begin{tabular}{@{}p{0.18\textwidth}p{0.50\textwidth}p{0.24\textwidth}@{}}
        \hline
        {Symbol} & {Meaning} & {Role} \\
        \hline
        {$\rvecx$} & {Stacked uncompressed sensor-mixture block} & {Physical observation} \\
        {$\widehat{\rvecx}_{\ratedist}$} & {Forward Gaussian RD test-channel output} & {Ideal benchmark observation} \\
        {$\widehat{\rvecx}$} & {Subtractive-dither reconstruction at the FC} & {Practical compressed observation} \\
        {$\rvecq,\matC_{\rvecq}$} & {Gaussian test-channel error and its covariance} & {RD benchmark distortion} \\
        {$\rvece,\matR_{\rvece}$} & {Dithered-quantization error and its sensor-domain covariance} & {Practical debiasing term} \\
        {$\matA,\matB$} & {Mixing matrix and true demixing matrix, $\matB=\matA^{-1}$} & {Unknown target and its inverse} \\
        {$\widehat{\rmatB}_N,\widehat{\rmatG}_N$} & {Estimated demixing matrix and gain matrix, $\widehat{\rmatG}_N=\widehat{\rmatB}_N\matA$} & {Separation output} \\
        {$\isr,\matC_{\crlb},\matC^{\rmatG}_{\icrlb}$} & {Interference-to-source ratio, mixing-matrix CRLB, and its transformation to the demixing and gain matrices} & {Performance metric and benchmark} \\
        \hline
    \end{tabular}
\end{table*}}

\vspace{-0.2cm}
\section{Identifiability Under Compressed Mixtures}\label{sec:identifiability}

This section addresses an important basic question that precedes any performance analysis: \emph{when is blind separation possible from compressed, distributed measurements?} We formalize separability directly in terms of the statistical model induced at the FC by the sensor-side compression mappings, and establish a data-processing principle showing that compression cannot improve separability relative to the uncompressed mixtures. We then specialize the discussion to the Gaussian ICA model and formulate conditions under which identifiability is preserved under compression.

\vspace{-0.3cm}
\subsection{Parameterization and ICA Ambiguities}\label{subsec:paramsambig}
Recall that even in the absence of compression, the ICA model admits intrinsic ambiguities. In particular, if $\matP\in\reals^{K\times K}$ is any permutation matrix and $\matD\in\reals^{K\times K}$ is any invertible diagonal matrix, then the re-parameterization
\begin{equation}
    \matA' \triangleq \matA\matP\matD,
    \quad
    \rvecs'[n] \triangleq \inv{\matD}\tp{\matP}\rvecs[n],
    \label{eq:ica_ambiguity}
\end{equation}
leaves the noiseless mixture $\matA\rvecs[n]$ unchanged. Accordingly, the best one can hope for, even in the noiseless case and all the more so in noisy mixtures, is to identify $\matA$ and $\rvecs[n]$ up to permutation and scaling. We formalize this by defining an equivalence relation $\sim$ on the space of invertible matrices that identifies mixing matrices that differ only by the ICA ambiguities. Concretely, $\matA\sim\matA'$ if there exist a permutation matrix $\matP$ and an invertible diagonal matrix $\matD$ such that
\begin{equation}
    \matA'=\matA\matP\matD .
    \label{eq:A_equiv_relation}
\end{equation}
Under this equivalent parameterization, the corresponding source vector is transformed according to $\rvecs'[n]=\inv{\matD}\tp{\matP}\rvecs[n]$, so that the recovered sources are meaningful only up to the same permutation and scaling ambiguity.

\subsection{Identifiability Under Compression}

Recall that $\rvecx\in\reals^{KN\times 1}$ denotes the stacked uncompressed observation block defined in~\eqref{eq:stacked_blocks_def}, and let $\Prob_{\theta}^{\rvecx}$ denote its probability law under parameter $\theta\in\Theta$ (e.g., the mixing matrix $\matA$). Since each sensor $\ell\in\setS_K$ applies a possibly randomized encoder to produce a message $\rndm_\ell\in\setM_\ell$ as in~\eqref{eq:encoder_general}, we model the encoder by a Markov kernel
\begin{align}
    &Q_\ell(\setB_\ell\mid \vecx_\ell)
    \triangleq
    \Prob\!\left(\rndm_\ell\in\setB_\ell \mid \rvecx_\ell=\vecx_\ell\right),
    \quad
    \setB_\ell\subseteq\setM_\ell.
    \label{eq:kernel_def}
\end{align}
Thus, for each realization $\vecx_\ell\in\reals^{N\times 1}$ of the local observation block, $Q_\ell(\cdot\mid\vecx_\ell)$ specifies the conditional distribution of the message generated by sensor $\ell$. We assume throughout that the encoders are fixed in advance and do \emph{not} depend on the unknown parameter $\theta$. We further assume the encoders operate \emph{independently} across sensors, i.e., there is no encoder cooperation or shared randomness. {When subtractive dithering is used, the same factorization holds conditionally on the collection of parameter-independent sensor--FC seeds.} Hence, the joint encoding law factors as a product kernel $Q(\cdot\mid \rvecx)\triangleq\prod_{\ell=1}^{K} Q_\ell(\cdot\mid \rvecx_\ell)$. 

The induced law of the tuple $\rndm\triangleq(\rndm_1,\ldots,\rndm_K)$ under $\theta$ is obtained by marginalizing the joint law over $\rvecx$, namely
\begin{equation}
    \Prob_{\theta}^{\rndm}(\setB)\!
    \triangleq \!
    \int_{\reals^{KN\times 1}}\! Q(\setB\mid \rvecx)\, \Prob_{\theta}^{\rvecx}({\rm d}\rvecx),
    \quad
    \setB\subseteq \setM_1\times\cdots\times\setM_K.
    \label{eq:pushforward_m}
\end{equation}
Therefore, after compression, all statistical information available for identifying the parameter is contained in the family of induced message laws $\{\Prob_{\theta}^{\rndm}\}_{\theta\in\Theta}$. This motivates the following identifiability notion, which mirrors the usual identifiability definition but accounts for the intrinsic ICA ambiguities.

\begin{definition}[Equidentifiability under compression]\label{def:identifiability_compressed}
The parameter $\theta\in\Theta$ is said to be \emph{equidentifiable}---identifiable up to the equivalence relation on $\Theta$---from the compressed messages $\rndm$ if for any $\theta,\theta'\in\Theta$,
\begin{equation}
    \Prob_{\theta}^{\rndm} = \Prob_{\theta'}^{\rndm}
    \quad
    \Longrightarrow
    \quad
    \theta'\sim \theta.
    \label{eq:ident_def_compressed}
\end{equation}
Equidentifiability from the uncompressed observations $\rvecx$ is defined analogously by replacing $\rndm$ with $\rvecx$.
\end{definition}

Definition~\ref{def:identifiability_compressed} is purely statistical and does not assume any particular separation algorithm, or any separation criterion. It captures the fundamental question of whether distinct parameter points induce distinct (up to $\sim$) distributions at the FC.

The following proposition formalizes the intuition that compression cannot improve equidentifiability. It is an equidentifiability analogue of the data processing inequality, and follows directly from the fact that $\Prob_{\theta}^{\rndm}$ is a pushforward of $\Prob_{\theta}^{\rvecx}$ through a parameter-independent kernel.

\begin{proposition}[Compression cannot improve equidentifiability]\label{prop:dpi_identifiability}
Assume that the encoder kernels $\{Q_\ell(\cdot\mid \rvecx_\ell)\}_{\ell\in\setS_K}$ do not depend on $\theta$. If there exist $\theta,\theta'\in\Theta$ such that $\theta'\not\sim\theta$ and
\begin{equation}
    \Prob_{\theta}^{\rvecx} = \Prob_{\theta'}^{\rvecx},
    \label{eq:nonident_x}
\end{equation}
then necessarily
\begin{equation}
    \Prob_{\theta}^{\rndm} = \Prob_{\theta'}^{\rndm}.
    \label{eq:nonident_m}
\end{equation}
In particular, non-equidentifiability from $\rvecx$ implies non-equidentifiability from $\rndm$, and equivalently, equidentifiability from $\rndm$ implies equidentifiability from $\rvecx$.
\end{proposition}

\begin{IEEEproof}
Fix any $\setB\subseteq \setM_1\times\cdots\times\setM_K$. By~\eqref{eq:pushforward_m} and~\eqref{eq:nonident_x},
\begin{align}
    \Prob_{\theta}^{\rndm}(\setB)
    &=
    \int Q(\setB\mid \rvecx)\, \Prob_{\theta}^{\rvecx}({\rm d}\rvecx)\\
    &=
    \int Q(\setB\mid \rvecx)\, \Prob_{\theta'}^{\rvecx}({\rm d}\rvecx)
    =
    \Prob_{\theta'}^{\rndm}(\setB).
\end{align}
Since $\setB$ is arbitrary, $\Prob_{\theta}^{\rndm}=\Prob_{\theta'}^{\rndm}$.
\end{IEEEproof}

\begin{remark}[Relation to data processing inequalities]\label{rem:dpi_divergence}
Proposition~\ref{prop:dpi_identifiability} can be viewed as the zero-divergence case of standard contraction results under Markov kernels. For example, for the Kullback--Leibler divergence $D(\cdot\|\cdot)$, from the data processing inequality,
\begin{equation}
    D\!\big(\Prob_{\theta}^{\rndm}\big\|\Prob_{\theta'}^{\rndm}\big)
    \le
    D\!\big(\Prob_{\theta}^{\rvecx}\big\|\Prob_{\theta'}^{\rvecx}\big),
\end{equation}
and similarly and more generally for $f$-divergences. Thus, if $\Prob_{\theta}^{\rvecx}=\Prob_{\theta'}^{\rvecx}$, then any such divergence between the induced message laws must also be zero, which implies~\eqref{eq:nonident_m}.
\end{remark}

Proposition~\ref{prop:dpi_identifiability} implies that compression can only \emph{preserve} or \emph{destroy} equidentifiability relative to the uncompressed model, but cannot create equidentifiability in cases where it is absent. {While this proposition is an elementary data-processing observation, it is included to anchor Definition~\ref{def:identifiability_compressed} and delimit what any compression rule can preserve. The substantive design question is not whether compression can create identifiability, but which encoder invariances destroy the original ICA equivalence classes and which compressed statistics remain sufficient for separation. The two compression examples analyzed later in Section~\ref{subsec:gausssourcesos} illustrate this distinction: sign compression is destructive, whereas subtractively dithered quantization can preserve source information sufficient for the relevant BSS equivalence class.}

\vspace{-0.3cm}
\section{Benchmarks and Notions of Separation Consistency Under Lossy Compression}\label{sec:benchmarks}

Lossy compression degrades the information available at the FC, so the classical notion of ``consistency'' should be formulated relative to an appropriate uncompressed benchmark. We next formalize an oracle-normalized notion of separation from compressed mixtures. Throughout, we consider signal models that are equidentifiable for uncompressed mixtures.

\vspace{-0.15cm}
\subsection{ISR Degradation Measures}

Let the FC produce a (data-dependent) demixing matrix $\widehat{\rmatB}_N \in\reals^{K\times K}$, with which we define the gain matrix $\widehat{\rmatG}_N \triangleq \widehat{\rmatB}_N \matA$. The component-wise ISR, defined in \eqref{eq:isr_k_def}, is denoted by $\isr_{k\ell}(\widehat{\rmatG}_N)$, and the average ISR, defined in \eqref{eq:isr_def}, is denoted by $\isr(\widehat{\rmatG}_N)$.
To handle the inherent ICA ambiguities, we henceforth fix any convenient ordering and scaling convention, without affecting the ISR-related results in the sequel.

When both the encoders and the decoder are part of the design, define the smallest achievable ISR under the rate constraints by
\begin{equation}\label{eq:isr_star_compressed}
    \isr_{N}^{\star}
    \triangleq
    \inf_{f,h}
    \isr\left(\widehat{\rmatG}_N\right),
\end{equation}
where the rate constraints are enforced via \eqref{eq:rate_constraint}, and for brevity we use $f$ and $h$ to denote all the encoders $\{f_{\ell}\}$ and all the decoders $\{h_{\ell}\}$, respectively. If the compression mappings are fixed \textit{a priori}, then $f$ is omitted from the infimum in~\eqref{eq:isr_star_compressed}. 

To quantify the degradation in separation quality due to compression, define the corresponding best achievable ISR when the FC has access to the \emph{uncompressed} mixture block $\rvecx$ (equivalently, encoder with infinite compression rate) as
\begin{equation}\label{eq:isr_star_uncompressed}
    \isr_{N,\infty}^{\star}
    \triangleq
    \inf_{ h_{\infty}}\ \isr\left(\widehat{\rmatG}_{N,\infty}\right),
\end{equation}
where $\widehat{\rmatB}_{N,\infty}\triangleq h_{\infty}(\rvecx)$ and $\widehat{\rmatG}_{N,\infty}\triangleq \widehat{\rmatB}_{N,\infty}\matA$, where $h_\infty(\cdot)$ is an operator that returns $\widehat{\matB}$, an estimate of the demixing matrix $\matB\triangleq\inv{\matA}$, based on the observed uncompressed data $\rvecx$. With \eqref{eq:isr_star_compressed} and \eqref{eq:isr_star_uncompressed}, we define the ISR degradation factor
\begin{equation}\label{eq:isr_ratio_def}
    \eta_N^{\isr}\triangleq \frac{\isr_{N}^{\star}}{\isr_{N,\infty}^{\star}},
\end{equation}
and the excess-ISR penalty
\begin{equation}\label{eq:isr_excess_def}
    \Delta_N^{\isr}\triangleq \isr_{N}^{\star}-\isr_{N,\infty}^{\star}.
\end{equation}
When $\isr_{N,\infty}^{\star}>0$ for all $N$, which is typical for finite-sample sizes and noisy models, the ratio~\eqref{eq:isr_ratio_def} is well-defined. We say that the system is \emph{ISR-oracle rate-consistent} if
\begin{equation}\label{eq:isr_oracle_consistency}
    \lim_{N\to\infty}\eta_N^{\isr}=\eta_\infty^{\isr}\in\positivereals,
\end{equation}
i.e., compression does not incur an asymptotic ISR penalty relative to the best uncompressed blind separation in terms of the convergence rate. More generally, $\eta_N^{\isr}\geq 1$ quantifies the nonasymptotic multiplicative ISR loss due to compression.
Alternatively, one may adopt the additive criterion \eqref{eq:isr_excess_def}. For this (somewhat weaker) measure, we say that the system is \emph{ISR-oracle consistent} if $\lim_{N\to\infty}\Delta_N^{\isr}=0$.

Consider the setting in which the compression rates of all the encoders are equal and the rate is denoted by $R$. An interesting and relevant special case is when for each fixed rate $R$, the compressed and uncompressed estimators of the demixing matrix are both first-order efficient, so that
\begin{align}
    \isr_{N}^{\star}(R)
    &=
    \frac{c_{\rm mix}(R)}{N}+o\!\left(N^{-1}\right),\label{eq:isr_asymptotic_constants1}\\
    \isr_{N,\infty}^{\star}
    &=
    \frac{c_{\rm mix}(\infty)}{N}+o\!\left(N^{-1}\right),\label{eq:isr_asymptotic_constants2}
\end{align}
for some positive constants $c_{\rm mix}(R), c_{\rm mix}(\infty)\in\positivereals$. Substituting \eqref{eq:isr_asymptotic_constants1} and \eqref{eq:isr_asymptotic_constants2} into \eqref{eq:isr_ratio_def} and \eqref{eq:isr_excess_def} yields
\begin{align}
    \eta_N^{\isr}(R)
    &=
    \frac{c_{\rm mix}(R)}{c_{\rm mix}(\infty)}+o(1)=\eta_\infty^{\isr}+o(1),
    \label{eq:isr_ratio_asymptotic}\\
    \Delta_N^{\isr}(R)
    &=
    \frac{c_{\rm mix}(R)-c_{\rm mix}(\infty)}{N}+o\!\left(N^{-1}\right).
    \label{eq:isr_excess_asymptotic}
\end{align}
Hence, beyond the more trivial conclusion that in this case the system is ISR-oracle consistent, the multiplicative compression penalty~\eqref{eq:isr_ratio_asymptotic} is asymptotically independent of $N$.

\vspace{-0.25cm}
\subsection{Gaussian Sources and SOS-based Separation}\label{subsec:gausssourcesos}

We now specialize to the Gaussian ICA model. For a block of length $N$, assume that $\rvecx \sim \normal(\veczero,\matC_{\rvecx})$, where $\matC_{\rvecx}$ is given by~\eqref{eq:Cx_finite_block}. In this case, the probability law of $\rvecx$ is fully determined by its covariance matrix. Consequently, any blind separation task that is based on $\rvecx$---and in particular, estimation of the mixing matrix $\matA$---can only depend on $\theta$ through $\matC_{\rvecx}(\theta)$. Equivalently, for Gaussian models, separability from $\rvecx$ is equivalent to separability from $\matC_{\rvecx}$,
\begin{equation}
    \Prob_{\theta}^{\rvecx} = \Prob_{\theta'}^{\rvecx}
    \quad
    \Longleftrightarrow
    \quad
    \matC_{\rvecx}(\theta)=\matC_{\rvecx}(\theta').
    \label{eq:gaussian_ident_cov}
\end{equation}
This motivates a SOS-based analysis: we next formulate conditions on the collection $\{\matC_{\rvecs_k}\}_{k\in\setS_K}$ under which the mixing matrix $\matA$ remains equidentifiable from the covariance structure of the observed mixtures.

In classical {noiseless} Gaussian ICA with {general} sources{ covariance
structures}, identifiability is enabled by \emph{second-order temporal diversity}{~\cite{yeredor2010blind}}. Intuitively, if the sources exhibit distinct covariance structure{ up to scaling}, then the collection of mixture covariances across {time indices} is associated with a unique demixing matrix, up to permutation and scale. In this setting, SOS diversity is captured by $\{\matC_{\rvecs_k}\}_{k\in\setS_K}$: if these matrices are {nonzero and pairwise nonproportional}, then $\matA$ is identifiable from $\matC_{\rvecx}$, even in the presence of sensor noise{ with known covariance}.\footnote{In the WSS specialization, the {analogous} condition {on the full autocovariance sequences }can be stated in terms of {pairwise nonproportional} power spectral densities (PSDs); see Section~\ref{sec:rd_model} for the stationary formulation used to obtain explicit rate-dependent characterizations.}

{More explicitly, pairwise nonproportionality requires
\begin{equation}
    \matC_{\rvecs_i}/\Tr(\matC_{\rvecs_i})
    \neq
    \matC_{\rvecs_j}/\Tr(\matC_{\rvecs_j}),
    \qquad
    \forall\, i\neq j\in\setS_K.
    \label{eq:source_cov_diversity}
\end{equation}
Equivalently, for each source pair, there is no scalar $\alpha>0$ such that $C_{\rnds_i}[n,m]\!=\!\alpha C_{\rnds_j}[n,m]$ for all $n,m\in\setS_N$. Under compression, \eqref{eq:source_cov_diversity} remains necessary but is not, by itself, sufficient for SOS-based
equidentifiability. The population SOS available at the FC must uniquely determine $\matA$ up to permutation and scaling under the source, noise, and compression models. A sufficient condition is that these SOS determine the uncompressed covariance $\matC_{\rvecx}$, provided that the uncompressed model is equidentifiable. In general, compression can introduce additional ambiguities even when \eqref{eq:source_cov_diversity} holds.}

Proposition~\ref{prop:dpi_identifiability} implies that equidentifiability under compression, i.e., of the model
$\{\Prob_{\theta}^{\rndm}\}_{\theta\in\Theta}$, is possible only if the uncompressed model
$\{\Prob_{\theta}^{\rvecx}\}_{\theta\in\Theta}$ is equidentifiable. In Gaussian ICA, separation is enabled by
SOS. Therefore, a compression rule must preserve sufficient correlation-based temporal diversity to retain equidentifiability, and in turn, enable source separation.

A simple sufficient condition is that the FC can exactly recover from the messages a statistic of the uncompressed mixtures that is itself equidentifying. Concretely, let $\vect:\reals^{KN{\times1}}\to\reals^{d{\times1}}$ be a deterministic statistic satisfying
\begin{equation}
    \Prob_{\theta}^{\vect(\rvecx)}
    =
    \Prob_{\theta'}^{\vect(\rvecx)}
    \quad\Longrightarrow\quad
    \theta'\sim\theta,
    \quad
    \forall\,\theta,\theta'\in\Theta .
    \label{eq:T_equidentifying}
\end{equation}
Suppose further that there exists a measurable mapping $\widehat{\vect}$ such that
\begin{equation}
    \Prob_{\theta}\!\left(
    \widehat{\vect}(\rndm)=\vect(\rvecx)
    \right)=1,
    \quad
    \forall\,\theta\in\Theta .
    \label{eq:T_recoverable_from_messages}
\end{equation}
Then $\theta$ is equidentifiable from $\rndm$.

In particular, for Gaussian ICA, if the recoverable statistic $\vect(\rvecx)$ consists of a SOS collection that uniquely determines the mixing matrix up to permutation and scaling (e.g., multiple ``views'' such as lagged or filtered covariances), then equidentifiability is preserved under compression. Conversely, if compression eliminates temporal diversity that distinguishes sources, equidentifiability may be lost, as demonstrated next.

We now illustrate both possibilities with two canonical examples: (i) a sign-only per-sensor compression rule that is invariant to positive row scalings of $\matA$, hence introduces new non-equidentifiability and prevents separation; and (ii) a subtractively dithered uniform quantizer (under a no-overload regime) that induces an additive, independent, zero-mean quantization noise with known variance, so that the mixture covariance can be exactly debiased and the SOS structure—and thus equidentifiability—can be preserved.

\begin{example}[Equidentifiability-destructive compression: sign quantization]\label{ex:equident_destructive_sign}
Consider a noiseless Gaussian mixture model with $K=2$ and invertible $\matA\in\reals^{2\times 2}$, i.e., $\rvecx[n]=\matA\rvecs[n]$, where $\rvecs[n]$ is a zero-mean non-degenerate Gaussian vector with independent components, and assume that the uncompressed model is equidentifiable via SOS temporal diversity.

Let each sensor transmit only the sign of its samples,
\begin{equation}\label{eq:ex_sign_compression_short}
    \rndm_\ell[n]\triangleq \sign(\rndx_\ell[n])\in\{-1,+1\},
    \;\;\; \ell\in\{1,2\}, \; n\in\integers,
\end{equation}
and let $\rvecm[n]\triangleq\tp{[\rndm_1[n]\ \rndm_2[n]]}$. Fix any positive row-scaling $\matD_{\rm row}=\diag(\alpha_1,\alpha_2)$ with $\alpha_1\neq \alpha_2\in\positivereals$, and define $\matA'\triangleq \matD_{\rm row}\matA$.
Then $\rvecx'[n]=\matA'\rvecs[n]=\matD_{\rm row}\rvecx[n]$, so $\rndx'_\ell[n]=\alpha_\ell\rndx_\ell[n]$ and hence
$\sign(\rndx'_\ell[n])=\sign(\rndx_\ell[n])$ for all $\ell,n$. Therefore,
\begin{equation}\label{eq:ex_sign_invariant_short}
    \rvecm'[n]\triangleq\tp{[\rndm'_1[n]\ \rndm'_2[n]]}=\rvecm[n]\qquad \forall\,n\in\integers,
\end{equation}
where $\{\rndm'_{\ell}[n]\}$ are as in \eqref{eq:ex_sign_compression_short}. It follows from \eqref{eq:ex_sign_invariant_short} that the message process has the same distribution under $\matA$ and $\matA'$.

However, $\matA'$ is generally \emph{not} equivalent to $\matA$ under the ICA ambiguity
$\matA\mapsto \matA\matP\matD$ (right multiplication by a permutation and diagonal scaling), since $\matA'$ is a \emph{left} diagonal scaling. Hence, the model is not equidentifiable from one-bit sign-quantization-type compression, and no procedure can separate the sources (even up to permutation and scaling) under \eqref{eq:ex_sign_compression_short}.
\end{example}

\begin{example}[Equidentifiability-preserving compression: subtractively dithered uniform quantization]\label{ex:equident_preserve_dither}
Consider again a mixture of Gaussian sources, where
$\rvecx\sim\normal(\veczero,\matC_{\rvecx})$ and $\theta$ is equidentifiable from the uncompressed mixtures. Fix $\Delta\in\positivereals$ and define the mid-tread uniform quantizer,
\begin{equation}\label{eq:ex_uniform_quantizer}
    \mathcal{Q}_\Delta(t)\triangleq \Delta\left\lfloor \frac{t}{\Delta}+\frac{1}{2}\right\rfloor,\qquad t\in\reals.
\end{equation}
Each sensor uses subtractive dithering: for every $\ell\in\setS_K$ and $n\in\setS_N$ draw
$\rndu_\ell[n]\sim\mathrm{Unif}(-\Delta/2,\Delta/2)$ i.i.d., independent of $\rvecx$, and assume $\{\rndu_\ell[n]\}$ are known
to the FC (e.g., via a shared seed). The sensor transmits $\mathcal{Q}_\Delta(\rndx_\ell[n]+\rndu_\ell[n])$, and the FC forms
\begin{equation}\label{eq:ex_subtractive_recon}
    \widehat{\rndx}_\ell[n]\triangleq \mathcal{Q}_\Delta(\rndx_\ell[n]+\rndu_\ell[n])-\rndu_\ell[n].
\end{equation}

To control overload, use a finite dynamic range $[-\mybar{x}_N,\mybar{x}_N]$ (with saturation outside), where $\mybar{x}_N$ is chosen so that
\begin{equation}\label{eq:ex_no_overload_event}
    \setE_N \triangleq \left\{\max_{\ell\in\setS_K, n\in\setS_N}|\rndx_\ell[n]+\rndu_\ell[n]|\le \mybar{x}_N\right\}
\end{equation}
satisfies $\Prob(\setE_N)\ge 1-\delta_N$ with $\delta_N\downarrow 0$ as $N\to\infty$. For Gaussian $\rvecx$, one may take
$\mybar{x}_N$ to scale as $\sqrt{\log(KN/\delta_N)}$ (up to constants), e.g., based on a union bound for Gaussian tails.

On the no-overload event $\setE_N$, no clipping occurs, and the
finite-range implementation agrees sample-wise with the standard
non-overloaded subtractively dithered uniform quantizer. For this standard
non-overloaded model, subtractive dithering yields the additive-noise
representation~\cite{gray1993dithered}
\begin{equation}\label{eq:ex_additive_model}
    \widehat{\rndx}_\ell[n]=\rndx_\ell[n]+\rnde_\ell[n],
    \quad
    \sigma_q^2\triangleq \Varop(\rnde_\ell[n])=\frac{\Delta^2}{12},
\end{equation}
where $\rnde_\ell[n]\perp\rvecx$ and $\{\rnde_\ell[n]\}$ are zero-mean i.i.d.\ across $(\ell,n)$. Denote by $\widehat{\rvecx}$ the stacked reconstruction and by $\rvece\triangleq\widehat{\rvecx}-\rvecx$ the stacked quantization error. Then, for the model~\eqref{eq:ex_additive_model},
\begin{equation}\label{eq:ex_cov_shift}
    \matC_{\widehat{\rvecx}}=\Covop(\widehat{\rvecx})
    =
    \Covop(\rvecx)+\Covop(\rvece)
    =
    \matC_{\rvecx}+\sigma_q^2\matI_{KN}.
\end{equation}
Since $\sigma_q^2$ is known, the FC can debias the mixture covariance via
\begin{equation}\label{eq:ex_cov_debias_short}
    \matC_{\rvecx}=\matC_{\widehat{\rvecx}}-\sigma_q^2\matI_{KN}.
\end{equation}
Therefore, any SOS-based equidentifiability condition that holds from
$\matC_{\rvecx}$ in the uncompressed Gaussian model continues to hold from
$\matC_{\widehat{\rvecx}}$ under the standard non-overloaded
dithered-quantization model. The finite-range implementation differs from
this model only on the overload event, whose probability is at most
$\delta_N$. Hence, this covariance-debiasing argument applies up to an
overload event with probability at most $\delta_N$.
\end{example}

{The two examples above also convey a practical design principle. Sign-only compression may remove sensor-amplitude information by making the compressed law invariant to positive row scalings of the mixing matrix, thereby creating ambiguities beyond the inevitable ICA permutation and source scaling. By contrast, subtractive dithering converts quantization, given no overload events, into an additive perturbation with known covariance. The fusion center can therefore subtract the zero-lag quantization contribution while preserving the source-dependent SOS diversity required for separation.}

\section{{Gaussian RD-based CRLB Benchmark} for Mixtures of Gaussian Sources}\label{sec:rd_model}

We now specialize the distributed compression-BSS model of Section~\ref{sec:problemformulation} to the Gaussian case and adopt an information-theoretic RD-optimal model for the sensor messages. {The Gaussian RD test channel is used here as an idealized, model-specific benchmark. It characterizes the separation information available after an MSE-optimal Gaussian reconstruction at the prescribed rate, but it is not a converse over all encoders satisfying the message-cardinality constraint, nor is it claimed to be attained by the practical scalar quantizer of Section~\ref{sec:practical_methods}. The subtractively dithered construction instead provides a realizable fixed-rate front end whose induced covariance and fourth-order cumulants can be corrected explicitly at the FC.} Throughout this section, we assume that for a fixed blocklength $N$, the stacked mixture block is zero-mean Gaussian,
\begin{equation}\label{eq:gauss_x_block}
    \rvecx \sim \normal(\veczero,\matC_{\rvecx}),
\end{equation}
with $\matC_{\rvecx}$ given by~\eqref{eq:Cx_finite_block}. We reiterate that the sensors do not cooperate, and each is subject to its own rate constraint~\eqref{eq:rate_constraint}.

\vspace{-0.25cm}
\subsection{Gaussian RD Test-Channel}\label{subsec:gauss_rd_test_channel}

The local observation block at sensor $\ell$ is Gaussian, i.e., $\rvecx_\ell \sim \normal(\veczero,\matC_{\rvecx_\ell})$, where $\matC_{\rvecx_\ell}\in\reals^{N\times N}$ is the $\ell$-th diagonal block of $\matC_{\rvecx}$. Under squared-error distortion, the per-sample rate constraint $R_\ell$ is captured by $I(\rvecx_\ell;\widehat{\rvecx}_{\ell,\ratedist}) \leq N R_\ell$, where $\widehat{\rvecx}_{\ell,\ratedist}$ denotes the forward Gaussian RD test-channel output available at the FC. The distortion associated with this test-channel is the normalized MMSE in estimating $\rvecx_\ell$ from $\widehat{\rvecx}_{\ell,\ratedist}$,
\begin{equation}\label{eq:dist_local_def}
    D_\ell \triangleq
    \frac{1}{N}
    \Exop\!\left[
    \left\|
    \rvecx_\ell
    -
    \Exop\!\left[\rvecx_\ell\mid \widehat{\rvecx}_{\ell,\ratedist}\right]
    \right\|_2^2
    \right].
\end{equation}

Let
$\matC_{\rvecx_\ell}\triangleq\matU_\ell \diag(\lambda_{\ell}[1],\ldots,\lambda_{\ell}[N])\tp{\matU_\ell}$
be an eigen-decomposition with $\lambda_{\ell}[1]\ge\cdots\ge\lambda_{\ell}[N]>0$, and define the Karhunen--Lo\`{e}ve transform (KLT)-domain vectors
\begin{equation}\label{eq:klt_def_local}
    \rvecy_\ell \triangleq \tp{\matU_\ell}\rvecx_\ell,
    \qquad
    \widehat{\rvecy}_{\ell,\ratedist}
    \triangleq
    \tp{\matU_\ell}\widehat{\rvecx}_{\ell,\ratedist},
\end{equation}
so that $\Covop(\rvecy_\ell)=\diag(\lambda_{\ell}[1],\ldots,\lambda_{\ell}[N])\in\positivereals^{N\times N}$. In the KLT domain, the forward Gaussian test-channel is
\begin{equation}\label{eq:rd_test_channel_local_y}
    \widehat{\rvecy}_{\ell,\ratedist}
    =
    \rvecy_\ell+\rvecq_{\ell,\rvecy},
    \quad
    \rvecq_{\ell,\rvecy}
    \sim
    \normal\!\left(
    \veczero,
    \diag(\nu_{\ell}[1],\ldots,\nu_{\ell}[N])
    \right),
\end{equation}
where $\rvecq_{\ell,\rvecy}$ is independent of $\rvecy_\ell$.

The Gaussian RD-optimal solution is characterized by reverse waterfilling~\cite{CoverThomas2006}: there exists a ``water level'' $\gamma_\ell\ge 0$ such that, for each eigenmode $n\in\setS_N$,
\begin{equation}\label{eq:reverse_waterfilling_d}
    d_{\ell}[n]
    \triangleq
    \Exop\!\left[
    \left(
    \rndy_{\ell}[n]
    -
    \Exop\!\left[
    \rndy_{\ell}[n]\mid\widehat{\rndy}_{\ell,\ratedist}[n]
    \right]
    \right)^2
    \right]
    =
    \min\{\lambda_{\ell}[n],\gamma_\ell\}.
\end{equation}
For active modes satisfying $\lambda_{\ell}[n]>d_{\ell}[n]$, the corresponding forward test-channel noise levels satisfy
\begin{equation}\label{eq:reverse_waterfilling_nu}
    \nu_{\ell}[n]
    =
    \frac{\lambda_{\ell}[n]\, d_{\ell}[n]}
    {\lambda_{\ell}[n]-d_{\ell}[n]}.
\end{equation}
If $d_{\ell}[n]=\lambda_{\ell}[n]$, the mode is discarded, the MMSE estimate of $\rndy_\ell[n]$ from $\widehat{\rndy}_{\ell,\ratedist}[n]$ is zero a.s., and the forward-noise representation may be interpreted as the limit $\nu_{\ell}[n]=+\infty$. The resulting per-sample rates and distortions are, respectively,
\begin{align}
    R_\ell
    &=
    \frac{1}{2}\cdot \frac{1}{N}
    \sum_{\{n\in\setS_N:\,\lambda_{\ell}[n]>\gamma_\ell\}}
    \log\!\left(\frac{\lambda_{\ell}[n]}{\gamma_\ell}\right),
    \label{eq:gauss_rd_rate_eigs}\\
    D_\ell
    &=
    \frac{1}{N}
    \sum_{n=1}^N \min\{\lambda_{\ell}[n],\gamma_\ell\}.
    \label{eq:gauss_rd_dist_eigs}
\end{align}

Equivalently, in the observation domain, the forward test-channel output is
\begin{equation}\label{eq:rd_test_channel_local}
    \widehat{\rvecx}_{\ell,\ratedist}
    =
    \rvecx_\ell+\rvecq_\ell,
    \qquad
    \rvecq_\ell\sim\normal(\veczero,\matC_{\rvecq_\ell}),
\end{equation}
where $\rvecq_\ell$ is independent of $\rvecx_\ell$, with
\begin{equation}\label{eq:Cq_eigform}
    \matC_{\rvecq_\ell}
    =
    \matU_\ell
    \diag(\nu_{\ell}[1],\ldots,\nu_{\ell}[N])
    \tp{\matU_\ell}.
\end{equation}
The FC then receives the stacked RD test-channel outputs
$\widehat{\rvecx}_{\ratedist}\triangleq
\tp{\left[\tp{\widehat{\rvecx}_{1,\ratedist}}\ \cdots\ \tp{\widehat{\rvecx}_{K,\ratedist}}\right]}$, which satisfy
\begin{equation}\label{eq:rd_test_channel_global}
    \widehat{\rvecx}_{\ratedist}
    =
    \rvecx+\rvecq
    =
    (\matA \otimes \matI_N)\rvecs+\rvecw+\rvecq,
    \quad
    \rvecq \triangleq \tp{\left[\tp{\rvecq_1}\ \cdots\ \tp{\rvecq_K}\right]}.
\end{equation}
Under independent per-sensor compression, $\rvecq_\ell\sim \normal(\veczero,\matC_{\rvecq_\ell})$ are independent across $\ell$ and independent of $\rvecx$. Hence, $\rvecq\sim\normal(\veczero,\matC_{\rvecq})$, where
\begin{equation}\label{eq:Cq_blockdiag}
    \matC_{\rvecq}
    =
    \blkdiag\{\matC_{\rvecq_1},\ldots,\matC_{\rvecq_K}\},
    \quad
    \matC_{\widehat{\rvecx}_{\ratedist}}
    =
    \matC_{\rvecx}
    +
    \matC_{\rvecq}.
\end{equation}

The forward Gaussian RD model \eqref{eq:rd_test_channel_local}--\eqref{eq:Cq_blockdiag} provides a tractable, closed-form parameterization of the statistical model of the compressed observations at the FC as a function of the rate budgets $\{R_\ell\}$ and the local covariances $\{\matC_{\rvecx_\ell}\}$. In what follows, we use this model to derive {a CRLB-based benchmark} for estimating $\matA$ and translate {it} into ISR bounds. {The resulting bound is conditional on the Gaussian RD test-channel model and its MSE-optimal sensor reconstructions. Directly optimizing an encoder for BSS or ISR would generally induce a different statistical experiment.}

\vspace{-0.25cm}
\subsection{{Gaussian RD-based }CRLB{ Benchmark} for Compressed ICA}
\label{subsec:crlb_compressed}

We consider estimation of the mixing matrix $\matA$ under the standard ICA
ambiguities of permutation and columnwise diagonal scaling. Since the mixing
matrix is identifiable only modulo these ambiguities, the CRLB below is
understood locally after fixing a representative of the corresponding
equivalence class. Specifically, we fix the local ordering of the sources
and impose an arbitrary column-normalization and sign convention on
$\matA$ in a neighborhood of the true mixing matrix. With this convention
fixed, $\vecop(\matA)$ denotes the vectorization of this fixed local
representative of the identifiable mixing matrix. The particular convention
is immaterial for the ISR bounds derived below, since the ISR criterion is
invariant to the standard ICA permutation and scaling ambiguities.

Let $\vectheta$ denote the finite-dimensional vector of unknown deterministic
parameters used in the CRLB derivation, including the entries of this fixed
representative of $\matA$, the sources' SOS parameters, and the sensor-noise
powers. Since $\widehat{\rvecx}_{\ratedist}$ is zero-mean Gaussian with covariance
\begin{align}\label{eq:compressed_covariance}
    \matC_{\widehat{\rvecx}_{\ratedist}} &= \matC_{\rvecx} + \matC_{\rvecq}\\
    &=(\matA\otimes\matI_N)\matC_{\rvecs}\tp{(\matA\otimes\matI_N)}
    + \matC_{\rvecw} + \matC_{\rvecq},
\end{align}
the Fisher information matrix for $\vectheta$ admits the standard Gaussian form
\begin{equation}\label{eq:fim_gaussian}
    \left[\matJ(\vectheta)\right]_{ij}
    =
    \frac{1}{2}
    \Tr\!\left(
        \matC_{\widehat{\rvecx}_{\ratedist}}^{-1}
        \frac{\partial\matC_{\widehat{\rvecx}_{\ratedist}}}{\partial \theta_i}
        \matC_{\widehat{\rvecx}_{\ratedist}}^{-1}
        \frac{\partial \matC_{\widehat{\rvecx}_{\ratedist}}}{\partial \theta_j}
    \right),
\end{equation}
where $\theta_i$ denotes the $i$-th element of $\vectheta$. {The dependence of the RD-optimal test channel on $\vectheta$ is
handled by considering a fixed compression design. Fix a nominal
operating point $\vectheta_0$ and use
$\matC_{\rvecx}(\vectheta_0)$ and the prescribed rates to construct the
KLTs, retained-mode sets, waterfilling levels, and forward Gaussian test
channels. Denote the resulting compression-noise covariance by
$\matC_{\rvecq,0}$. These design quantities are then held fixed while
$\vectheta$ varies in the likelihood used for estimation. Consequently,
on the retained test-channel subspace,
\begin{equation}\label{eq:fixed_design_covariance}
\matC_{\widehat{\rvecx}_{\ratedist}}(\vectheta)
=
\matC_{\rvecx}(\vectheta)+\matC_{\rvecq,0},
\qquad
\frac{\partial
\matC_{\widehat{\rvecx}_{\ratedist}}(\vectheta)}
{\partial\theta_i}
=
\frac{\partial\matC_{\rvecx}(\vectheta)}
{\partial\theta_i},
\end{equation}
since $\partial\matC_{\rvecq,0}/\partial\theta_i=\matzero$. At
$\vectheta_0$, this fixed front end coincides with the Gaussian
RD-optimal design at the prescribed rates, and it need not remain RD-optimal
away from $\vectheta_0$. Redesigning the compression mechanism as
$\vectheta$ varies would instead require differentiating the resulting
parameter-dependent test-channel covariance and would define a different
benchmark.} Thus,{ as reflected from \eqref{eq:compressed_covariance},} relative to the uncompressed model,
compression degrades the Fisher information solely through the additive
compression distortion covariance $\matC_{\rvecq}$.

We treat the sources' SOS and the sensor-noise powers as nuisance parameters and seek the CRLB for $\matA$. For any symmetric $\matX\in\reals^{N\times N}$, let $\operatorname{svec}(\matX)\in\reals^{N(N+1)/2}$ collect the upper-triangular
entries, including the diagonal. Let $\matD_N$ be the duplication matrix so that
\begin{equation}
    \vecop(\matX)=\matD_N\operatorname{svec}(\matX),
    \quad
    \operatorname{svec}(\matX)=\matD_N^{\dagger}\vecop(\matX),
\end{equation}
where $\matD_N^{\dagger}$ denotes the Moore--Penrose inverse. We denote
\begin{equation}
    \vectheta
    \triangleq
    \tp{\left[\tp{\vectheta_{\matA}} \; \tp{\vectheta_{\matA^{\rm\!C}}}\right]},
    \qquad
    \vectheta_{\matA}
    \triangleq
    \vecop(\matA)\in\reals^{K^2\times 1},
\end{equation}
where $\matA$ is understood according to the local convention fixed above, and
\begin{align}
    \vectheta_{\matA^{\rm\!C}}
    &\triangleq
    \tp{\left[
    \tp{\operatorname{svec}(\matC_{\rvecs_1})}
    \; \cdots \;
    \tp{\operatorname{svec}(\matC_{\rvecs_K})}
    \; \tp{\vecsigma}_{\rndw}
    \right]},\\
    \vecsigma_\rndw
    &\triangleq
    \tp{[\sigma_{\rndw,1}^2\ \cdots\ \sigma_{\rndw,K}^2]}
    \in\reals^{K\times 1}.
\end{align}
With the local convention fixed, we retain the ambient notation
$\vecop(\matA)\in\reals^{K^2\times 1}$ for notational simplicity; the CRLB is
understood on the corresponding identifiable local representative.

With $\matE_{\ell k}$ denoting the $K\times K$ elementary matrix, we have
\begin{align}\label{eq:dC_da_struct}
    \frac{\partial \matC_{\widehat{\rvecx}_{\ratedist}}}{\partial a_{\ell k}}
    &=
    (\matE_{\ell k}\otimes\matI_N)\matC_{\rvecs}(\tp{\matA}\otimes\matI_N)\\
    &+
    (\matA\otimes\matI_N)\matC_{\rvecs}(\tp{\matE}_{\ell k}\otimes\matI_N).
\end{align}
For an entry in $\operatorname{svec}(\matC_{\rvecs_k})$, say the $(u,v)$ upper-triangular element with $1\le u\le v\le N$, we also have
\begin{equation}\label{eq:dC_dCs_struct_entry}
    \frac{\partial \matC_{\widehat{\rvecx}_{\ratedist}}}{\partial [\matC_{\rvecs_k}]_{uv}}
    =
    (\matA\matE_{kk}\tp{\matA})\otimes
    \left(
    \matE_{uv}+\matE_{vu}-\mathbbm{1}_{\{u=v\}}\matE_{uv}
    \right).
\end{equation}
Additionally, since $\matC_{\rvecw}=\blkdiag\{\sigma_{\rndw,1}^2\matI_N,\ldots,\sigma_{\rndw,K}^2\matI_N\}$,
\begin{equation}\label{eq:dC_dsigw}
    \frac{\partial \matC_{\widehat{\rvecx}_{\ratedist}}}{\partial \sigma_{\rndw,\ell}^2}
    =
    \matE_{\ell\ell}\otimes \matI_N,
    \quad \ell\in\setS_K.
\end{equation}

Proceeding, we define the blockwise partition
\begin{equation}
    \matJ(\vectheta)\triangleq
    \begin{bmatrix}
        \matJ_{\matA\matA} & \matJ_{\matA{\matA^{\rm\!C}}}\\
        \matJ_{{\matA^{\rm\!C}}\matA} & \matJ_{{\matA^{\rm\!C}}{\matA^{\rm\!C}}}
    \end{bmatrix},
\end{equation}
where every entry is computed by~\eqref{eq:fim_gaussian} using \eqref{eq:dC_da_struct}--\eqref{eq:dC_dsigw}.
Under the above local convention, the CRLB for $\vectheta_{\matA}$ {for estimators that are locally unbiased at the nominal operating point} is given by
\begin{equation}\label{eq:crlb_A_schur_struct}
    \Covop\!\left(\widehat{\vectheta}_{\matA}\right)
    \succeq
    \left(
        \matJ_{\matA\matA}
        -
        \matJ_{\matA{\matA^{\rm\!C}}}
        \matJ_{{\matA^{\rm\!C}}{\matA^{\rm\!C}}}^{-1}
        \matJ_{{\matA^{\rm\!C}}\matA}
    \right)^{-1}
    \triangleq
    \matC_{\crlb}.
\end{equation}

The Schur complement~\eqref{eq:crlb_A_schur_struct} becomes analytically tractable once the nuisance SOS are given a low-dimensional structure, such as a Toeplitz model parameterized by ${T_{\rm lag}}+1$ correlation lags, where each $\matC_{\rvecs_k}$ is determined by its respective ${T_{\rm lag}}+1$ correlation parameters, or by a spectral parameterization on a fixed frequency grid. In these cases, $\matJ_{{\matA^{\rm\!C}}{\matA^{\rm\!C}}}$ has dimension $O(K{T_{\rm lag}})$, rather than $O(KN^2)$, and the Schur complement can often
be simplified, sometimes even diagonalized, in the same eigenbasis used by the RD test-channel, leading to semi-closed-form
expressions for the CRLBs of the entries of $\matA$.

{To obtain the CRLB for the demixing matrix
$\matB=\matA^{-1}$, let $\widehat{\rmatB}_N$ be any finite-covariance estimator that is
locally unbiased for $\matB=\matA^{-1}$ at the nominal operating point, in the standard CRLB sense with respect to the full parameter vector $\vectheta$. The Jacobian of the inverse map is
\begin{equation}\label{eq:inverse_map_jacobian}
    \matH_{\matB}
    \triangleq
    \frac{\partial\vecop(\matB)}{\partial\tp{\vecop(\matA)}}
    =
    -(\tp{\matB}\otimes\matB).
\end{equation}
Since $\matB$ does not depend on the nuisance parameters, the CRLB transformation rule for a differentiable function of $\vectheta$ gives
\begin{equation}\label{eq:crlb_B_transform}
    \Covop\!\left(\vecop\!\left(\widehat{\rmatB}_N\right)\right)
    \succeq
    \matH_{\matB}\matC_{\crlb}\tp{\matH_{\matB}}.
\end{equation}

Define $\widehat{\rmatG}_N\triangleq\widehat{\rmatB}_N\matA$. Since $\matA$ is fixed at its true value when evaluating separation performance,
\begin{align}\label{eq:crlb_G_transform}
    \Covop\!\left(\vecop\!\left(\widehat{\rmatG}_N\right)\right)
    &=(\tp{\matA}\otimes\matI_K)
    \Covop\!\left(\vecop\!\left(\widehat{\rmatB}_N\right)\right)
    (\matA\otimes\matI_K)\\
    &\succeq
    (\matI_K\otimes\matB)\matC_{\crlb}
    (\matI_K\otimes\tp{\matB})
    \triangleq \matC^{\rmatG}_{\icrlb},
\end{align}
}where $\matC^{\rmatG}_{\icrlb}$ is the CRLB propagated on the gain perturbations, or simply the induced CRLB (iCRLB).
{Here, we used $\matB\matA=\matI_K$. Local unbiasedness of $\widehat{\rmatB}_N$ implies $\Exop[\widehat{\rmatG}_N]=\matI_K$ at the nominal point. Hence, for every $k\neq\ell$,
\begin{equation}\label{eq:gkl_exact_crlb}
    \Exop\!\left[\widehat\rndg_{k\ell}^{\,2}\right]
    =\Varop\!\left(\widehat\rndg_{k\ell}\right)
    \ge
    \tp{\vece_{k\ell}}\matC^{\rmatG}_{\icrlb}\vece_{k\ell},
\end{equation}
}where $\vece_{k\ell}\in\reals^{K^2\times 1}$ denotes the canonical basis vector such that, for any $\matG\in\reals^{K\times K}$, $[\matG]_{k\ell}=\tp{\vece_{k\ell}}\vecop(\matG)$.
Recalling \eqref{eq:isr_k_def}, we obtain the iCRLB on the ISR, so that for any pair $k\neq \ell$,
\begin{align}
    \isr_{k\ell}\!\left(\widehat{\rmatG}_N\right)
    &\ge
    \left[
    \tp{\vece_{k\ell}}
    \matC^{\rmatG}_{\icrlb}
    \vece_{k\ell}\right]\cdot \frac{\Tr(\matC_{\rvecs_\ell})}{\Tr(\matC_{\rvecs_k})},
    \label{eq:isr_lb_from_crlb_clean}
\end{align}
{Thus, under the specified Gaussian RD test-channel model, the resulting expression is a CRLB on the ISR for locally unbiased estimators of the demixing matrix.}

\section{Practical Compression--Separation Methods}
\label{sec:practical_methods}

Having discussed the theoretical aspects related to separability under compression in distributed BSS, we now propose two practical compression--separation schemes. In both cases considered here, the compression rule is chosen so that the induced statistical model at the FC is explicit and tractable, and the separation stage is adapted to that modified model.

Throughout, sensors apply fixed-rate subtractively dithered uniform quantization, which induces a precise additive-noise model at the FC (on a high-probability no-overload event, to be defined below). Separation is then performed by using criteria that are matched to the induced statistical model.

{The Gaussian test channel of Section~\ref{sec:rd_model} and the dithered scalar quantizer therefore serve different purposes. The former is an ideal reconstruction benchmark with a closed-form covariance model. The latter is an implementable compression rule with an exact no-overloaded additive-noise representation. Thus, the practical methods below should not be expected to attain the Gaussian RD-based CRLB. Rather, their role is to demonstrate the gain from matching the separation statistics to the distortion induced by a realizable quantizer.}

\vspace{-0.25cm}
\subsection{Compression Front-End}\label{subsec:compression_frontend}

Fix per-sensor rate budgets $\{R_\ell\}_{\ell\in\setS_K}$, where for this practical scalar quantization scheme, we take $R_\ell\in\naturals$ bits/sample. Let $M_\ell \triangleq 2^{R_\ell}$ and choose a per-sensor symmetric dynamic range $\mybar{x}_\ell>0$. Define the uniform step size
\begin{equation}
    \Delta_\ell \triangleq \frac{2\mybar{x}_\ell}{M_\ell}, \quad \ell\in\setS_K.
\end{equation}
For each $\ell\in\setS_K$ and $n\in\setS_N$:
\begin{enumerate}
\item \textbf{Dithering:} Draw subtractive dither
\begin{equation}
    \rndu_\ell[n]
    \sim \mathrm{Unif}\!\left(-\frac{\Delta_\ell}{2},
                               \frac{\Delta_\ell}{2}\right),
\end{equation}
independently across $(\ell,n)$ and of the mixtures.

\item \textbf{Clipping:} Form the clipped input
\begin{equation}
    \rndz_\ell[n]
    \triangleq
    \max\{-\mybar{x}_\ell,
         \min\{\rndx_\ell[n]+\rndu_\ell[n],\mybar{x}_\ell\}\}.
\end{equation}

\item \textbf{Quantization:} Transmit the fixed-rate quantizer output
\begin{align}
    \rndm_\ell[n]
    \triangleq
    \mathcal{Q}_{\Delta_\ell}(\rndz_\ell[n]),
\end{align}
where $\mathcal{Q}_{\Delta_\ell}(\cdot)\!:\![ -\mybar{x}_\ell,\mybar{x}_\ell ]\!\to\!\{
    -\mybar{x}_\ell-\frac{\Delta_\ell}{2}
    + j\Delta_\ell
     \mid j\in\setS_{M_\ell}  \}$.
\end{enumerate}

At the FC, compute
\begin{equation}
    \widehat{\rndx}_\ell[n]
    \triangleq
    \rndm_\ell[n]-\rndu_\ell[n].
\end{equation}
Define the ``no-overload" event
\begin{equation}\label{eq:nooverloaddef}
    \setE_{N,\ell}
    \triangleq
    \left\{
    \max_{n\in\setS_N}
    |\rndx_\ell[n]+\rndu_\ell[n]|
    \le \mybar{x}_\ell
    \right\}.
\end{equation}
On the no-overload event $\setE_N \triangleq \cap_{\ell\in\setS_K}\setE_{N,\ell}$, no clipping occurs, and the quantizer operates in the standard
non-overloaded subtractive-dithering regime. In this regime, subtractive dithering yields the additive-noise model of \eqref{eq:ex_additive_model}, with $\Delta$ replaced by the corresponding sensor-dependent step size $\Delta_\ell$. Stacking across sensors, we have
\begin{align}
    \widehat{\rvecx}[n]  = \rvecx[n] + \rvece[n] &= \matA\rvecs[n] + \rvecw[n] + \rvece[n] \label{eq:modifiedmodelV0}\\
    &\triangleq \matA\rvecs[n] + \widetilde{\rvecw}[n].\label{eq:modifiedmodel}
\end{align}
Thus, under this additive-noise model, the FC observes the modified statistical model \eqref{eq:modifiedmodel} with effective noise $\widetilde{\rvecw}[n]=\rvecw[n]+\rvece[n]$, whose covariance matrix is given by
\begin{align}\label{eq:effectivenoisecovariance}
    \matR_{\widetilde{\rvecw}}
    &\triangleq \Exop\left[\widetilde{\rvecw}[n]\tp{\widetilde{\rvecw}[n]}\right] \\
    &=\Exop\left[\rvecw[n]\tp{\rvecw[n]}\right] + \Exop\left[\rvece[n]\tp{\rvece[n]}\right]
    \triangleq \matR_{\rvecw} + \matR_{\rvece},
\end{align}
where
\begin{equation}
    \matR_{\rvece} \triangleq \diag\!\left(\frac{\Delta_1^2}{12},\ldots,\frac{\Delta_K^2}{12}\right)\in\reals^{K\times K},
\end{equation}
and we emphasize that $\matR_{\rvece}$ is, by design, known to the FC.

{The dither sequence is generated from a pseudorandom seed shared by
each sensor and the FC. This shared-randomness resource is assumed
available, and its amortized seed cost is not included in $R_\ell$. The
dynamic range is selected as a function of the sensor scale and the
blocklength to control the probability of the no-overload event defined
in \eqref{eq:nooverloaddef}. As discussed in Section~\ref{subsec:gausssourcesos}, for Gaussian observations, logarithmic
growth of the dynamic range with the blocklength is sufficient under a
Gaussian tail bound. Debiasing removes the known quantization contribution
at the population level but does not remove finite-sample variance. Thus,
at very low rates or short blocklengths, the corrected covariance and
cumulant estimates can remain noisy even when their bias is removed. The
corrected covariance is therefore symmetrized and, when necessary,
eigenvalue-regularized before whitening.}

{The practical front end is intentionally a transparent fixed-rate quantizer and is not intended to model an application-specific codec. Given no overloads, dithered quantization provides a known additive-error distribution, permitting controlled analysis and correction of the second- and fourth-order statistics used for separation. Extending the analysis to transform, learned, or perceptual codecs requires a model of their induced multivariate statistics and is left for future work.}

Both methods described next employ the above sensor-side compression, and only the FC-side separation stage differs. 

\subsection{ICA of i.i.d.\ Sources with Cumulant Debiasing}
\label{subsec:method_iid_corrected}

In the first setting, we assume that sources are i.i.d.\ across time ($n$), mutually independent, and at most one source is Gaussian. The separation algorithm in this case is {given in Algorithm~1.}
\begin{figure}[!t]
\centering
\tcbset{colframe=black!90!blue,size=small,width=0.49\textwidth,halign=flush center,arc=2mm,outer arc=1mm}
\begin{tcolorbox}[upperbox=visible,colback=white,halign=left]
	\textbf{{Algorithm 1: }Cumulant-based Separation of Compressed Mixtures}\\
	\begin{enumerate}

\item Form the compressed batch $\widehat{\rmatX}\triangleq[\widehat{\rvecx}[1]\ \cdots\ \widehat{\rvecx}[N]]$;

\item Compute the empirical mean $\widehat{\rvecmu}\triangleq\frac{1}{N}\sum_{n=1}^N \widehat{\rvecx}[n]$;

\item Form the centered batch $\widetilde{\rmatX}\triangleq[\widehat{\rvecx}[1]-\widehat{\rvecmu}\ \cdots\  \widehat{\rvecx}[N]-\widehat{\rvecmu} ]$;

\item Compute the sample covariance $\widehat{\rmatR}_{\widehat \rvecx} = \frac{1}{N}\widetilde{\rmatX}\tp{\widetilde{\rmatX}}${ and its debiased version $\widehat{\rmatR}_{\rvecx}\triangleq\widehat{\rmatR}_{\widehat \rvecx}-\matR_{\rvece}$};

\item Whiten the centered batch: Choose any whitening matrix $\rmatV\in\reals^{K\times K}$ such that $\rmatV{\widehat{\rmatR}_{\rvecx}}\tp{\rmatV}=\matI_K$, and define the whitened data matrix $\rmatZ\triangleq \rmatV\widetilde{\rmatX}$;

\item Estimate $\setC_4(\rvecz[n])$, the fourth-order cumulant tensor of $\rvecz[n]$, based on $\rmatZ$ (recall that the columns of $\rmatZ$ are i.i.d.), and denote this estimator as $\widehat{\setC}_4(\rvecz)$;

\item Debias cumulants using the known $\setC_4(\rvece[n])$, which is diagonal and computable from $\{\Delta_\ell\}$;

\item Apply a cumulant-based joint diagonalization to the debiased tensor (e.g., JADE~\cite{cardoso1993blind}), and denote its output as $\widehat{\rmatU}$; and

\item Output $\widehat{\rmatB} \triangleq \widehat{\rmatU}\rmatV.$

\end{enumerate}
\end{tcolorbox}\vspace{-0.5cm}
\end{figure}
For completeness, we write explicitly the debiasing step.
For indices $i,j,k,\ell\in\setS_K$, define
\begin{equation}
\big[\widehat{\setC}_4^{\,\mathrm{debiased}}\big]_{i j k \ell}
=
\big[\widehat{\setC}_4(\rvecz)\big]_{i j k \ell}
-
\big[\setC_4(\rmatV\rvece)\big]_{i j k \ell}.
\end{equation}
Since the components of $\rvece[n]$ are independent,
\begin{equation}
\big[\setC_4(\rmatV\rvece)\big]_{i j k \ell} = \sum_{p=1}^K
v_{i p}v_{j p}v_{k p}v_{\ell p}
\,
\kappa_4(\rnde_p),
\end{equation}
where $\kappa_4(\rnde_p)=-\frac{\Delta_p^4}{120}$ denotes the fourth-order cumulant
of the $p$-th component of $\rvece[n]$, which is fully determined
by $\Delta_p$.

Under the additive-noise model \eqref{eq:modifiedmodelV0}, cumulants are
additive before whitening, namely,
\begin{equation}
\setC_4\!\left(\widehat{\rvecx}[n]\right)
=
\setC_4\!\left(\rvecx[n]\right)
+
\setC_4\!\left(\rvece[n]\right),
\end{equation}
since $\rvecx[n]$ and $\rvece[n]$ are independent. Consequently, for any
fixed matrix $\matV$,
\begin{equation}
\setC_4\!\left(\matV\widehat{\rvecx}[n]\right)
=
\setC_4\!\left(\matV\rvecx[n]\right)
+
\setC_4\!\left(\matV\rvece[n]\right).
\end{equation}
In the finite-sample algorithm, the whitening matrix $\rmatV$ is estimated from the compressed data {through the debiased covariance} and is therefore data-dependent. We use the above identity as a plug-in debiasing rule with the realized whitening matrix $\rmatV${, which is used both for whitening and for computing the transformed uniform-noise cumulant correction}.

Since $\rvece[n]$ has independent components with known marginal laws
determined by $\{\Delta_\ell\}$, the correction term
$\setC_4(\rmatV\rvece[n])$ is computable from the realized whitening matrix
$\rmatV$ and $\{\Delta_\ell\}_{\ell\in\setS_K}$, and can therefore be subtracted
explicitly.

The finite-range quantizer deviates from the additive-noise model \eqref{eq:modifiedmodelV0} only on
the overload event $\setE_N^c$. Since $\Prob(\setE_N)\ge 1-\delta_N$ with
$\delta_N\xrightarrow[]{N\to\infty}0$ for appropriate scaling of
$\mybar{x}_\ell$, the additive-noise model holds with probability tending
to one.

\vspace{-0.2cm}
\subsection{SOS-Based Separation with Covariance Debiasing}
\label{subsec:method_sos_ml_corrected}

In the second setting, we assume that sources are zero-mean WSS with unknown but distinct SOS. Under the compression described in Section~\ref{subsec:compression_frontend}, \eqref{eq:modifiedmodel} is the modified signal model with effective noise covariance \eqref{eq:effectivenoisecovariance}. Hence, the temporal covariance matrices of the compressed mixtures are given by
\begin{align}\label{eq:biasedsos}
    \hspace{-0.05cm}\matR_{\widehat \rvecx}[\tau]&\triangleq \Exop\left[\widehat\rvecx[n]\tp{\widehat\rvecx[n-\tau]}\right] =\begin{cases}
        \matR_{\rvecx}[0] + \matR_{\rvece}, & \tau=0\\
        \matR_{\rvecx}[\tau], & \tau\neq 0
    \end{cases},
\end{align}
where $\matR_{\rvecx}[\tau]\triangleq\Exop\left[\rvecx[n]\tp{\rvecx[n-\tau]}\right]$.

The SOS-based separation in this case is {given in Algorithm~2.}
\begin{figure}[!t]
    \centering
\tcbset{colframe=black!90!blue,size=small,width=0.49\textwidth,halign=flush center,arc=2mm,outer arc=1mm}
\begin{tcolorbox}[upperbox=visible,colback=white,halign=left]
	\textbf{{Algorithm 2: }SOS-based Separation of Compressed Mixtures}\\
\begin{enumerate}
\item Estimate the temporal covariances of the mixtures:
\begin{equation}
    \widehat{\matR}_{\widehat \rvecx}[\tau]
    =
    \frac{1}{N-\tau}
    \sum_{n=1}^{N-\tau}
    \widehat{\rvecx}[n+\tau]
    \tp{\widehat{\rvecx}[n]}.
\end{equation}

    \item Debias the zero-lag covariance:
\begin{equation}
    \widehat{\matR}_{\rvecx}[0]
    \triangleq
    \widehat{\matR}_{\widehat \rvecx}[0]
    -
    \matR_{\rvece},
    \end{equation}
    and for $\tau\neq 0$, set
    \begin{equation}
    \widehat{\matR}_{\rvecx}[\tau]
    \triangleq
    \widehat{\matR}_{\widehat \rvecx}[\tau].
    \end{equation}

\item Apply any SOS-based separation algorithm to the covariances $\{\widehat{\matR}_{\rvecx}[\tau]\}$ (e.g., SOBI~\cite{Belouchrani1997SOBI}); and

\item Output $\widehat{\rmatB}=\widehat{\rmatA}^{-1}$.
\end{enumerate}
\end{tcolorbox}
\end{figure}
Recall that in this case, all sources may be Gaussian and still separable~\cite{yeredor2010blind}. Accordingly, one may alternatively adopt the Gaussian ML framework~\cite{weiss2019maximum}, to be used here as a QML criterion~\cite{weiss2018performance}, which also accounts for unknown (distinct) temporal covariance matrices.

As shown in \eqref{eq:biasedsos}, compression via fixed-rate subtractively dithered uniform quantization alters only the effective noise covariance in the signal model. After subtracting $\matR_{\rvece}$ from the zero-lag estimate, the second-order structure matches that of the uncompressed mixtures, with modified noise level. Therefore, identifiability is governed by the same SOS diversity conditions as in the uncompressed ML framework. As long as compression preserves equidentifiability of the induced model, the consistent SOS-based estimators remain consistent.

\vspace{-0.15cm}
\section{Simulation Results}\label{sec:sim_results}

In this section, we report three numerical simulation studies that complement the theory of Sections~\ref{sec:rd_model} and~\ref{sec:practical_methods}: a Gaussian WSS benchmark under the RD test-channel, a practical finite-rate RF experiment with non-Gaussian sources, and a phase-transition study for distributed industrial acoustic monitoring with both Gaussian surrogates and real machine sounds. For simplicity, throughout we set $\sigma_{\rndw_{\ell}}^2=\sigma_{\rndw}^2$ for all $\ell\in\setS_K$, and the sources are normalized to have unit average power.

{Consistent with the non-Bayesian formulation, each mixing matrix is treated as an unknown deterministic parameter. In each experiment, a full-rank, moderately conditioned matrix was generated once at random and then held fixed across the Monte Carlo (MC) trials, while the source, observation-noise, and compression realizations were regenerated. The resulting averages therefore estimate performance for a fixed mixing matrix and permit the effects of rate, blocklength, source statistics, and debiasing to be examined without variation in the mixing matrix. The three experiments use different fixed matrices, and additional independently generated fixed matrices exhibited the same qualitative trends.}

\vspace{-0.35cm}
\subsection{Gaussian WSS Sources Under RD Compression}

We first consider the Gaussian SOS setting of Sections~\ref{subsec:gauss_rd_test_channel}--\ref{sec:practical_methods}. There are $K=L=3$ sources and sensors, and a randomly chosen full-rank mixing matrix, fixed throughout all trials,
\begin{equation}
\matA=\begin{bmatrix}
1.00 & 0.32 & -0.14\\
0.21 & 0.95 & 0.28\\
-0.10 & 0.24 & 1.08
\end{bmatrix}.
\end{equation}
The source processes are zero-mean Gaussian and WSS, with distinct spectra produced by the three shaping models summarized in Table~\ref{tab:exp1_sources}. Additive sensor noise is white Gaussian with observation signal-to-noise ratio (SNR) equal to $20$~dB,\footnote{Since the sources are normalized to have unit average power, throughout, the SNR is defined as the ratio between the average row power of $\matA$ and $\sigma_{\rndw}^2.$} and each sensor is compressed independently according to the Gaussian RD test-channel with rates $R\in\{1,1.5,2,\ldots,10\}$.{ For each $(N,R)$ point, the RD test-channel noise levels and retained-mode sets are constructed once from the nominal sensor spectra and are then held fixed in both the ML estimator and the FIM
calculation.}

\begin{table}[t]
    \centering
    \caption{Experiment~1 source models. Normalized frequencies are in cycles/sample, and each process is scaled to unit variance.}
    \label{tab:exp1_sources}
    \footnotesize
    \setlength{\tabcolsep}{3.5pt}
    \renewcommand{\arraystretch}{1.05}
    \begin{tabular}{@{}c l p{0.54\linewidth}@{}}
        \hline
        Source & Model & Parameters \\
        \hline
        1 & Lowpass FIR & $121$ taps; cutoff $0.09$. \\
        2 & Bandpass FIR & $121$ taps; passband $[0.20,\,0.30]$. \\
        3 & Narrowband AR(2) & Poles at $0.92e^{\pm \jmath2\pi\cdot 0.40}$. \\
        \hline
    \end{tabular}
\end{table}

This experiment is intended as an oracle-aided benchmark for validating
the RD-induced CRLB and the predicted rate dependence, rather than as a fully
blind implementation. Accordingly, the FC applies the Gaussian ML
estimator~\cite{weiss2019maximum} for the mixing matrix using the known source
PSDs and the compressed Gaussian model. To isolate statistical performance
from global optimization effects, we implement the estimator via Fisher
scoring initialized at the true mixing matrix.\footnote{The initialization at the true mixing matrix is used only to study the local statistical behavior of the estimator. For any finite sample size $N$, the ML
estimate is almost surely distinct from this initialization.} We report results for three sample sizes, namely
$N\in\{100,200,500\}$, and average over $1000$ MC trials per
$(N,R)$ point.

{The true-$\matA$ initialization deliberately removes the global optimization question from this experiment. Consequently, agreement with the iCRLB demonstrates local statistical efficiency of the oracle-aided likelihood implementation only. Establishing whether a fully blind, realizable algorithm can reliably approach this local benchmark is a separate algorithmic question and is outside the scope of the present work.}

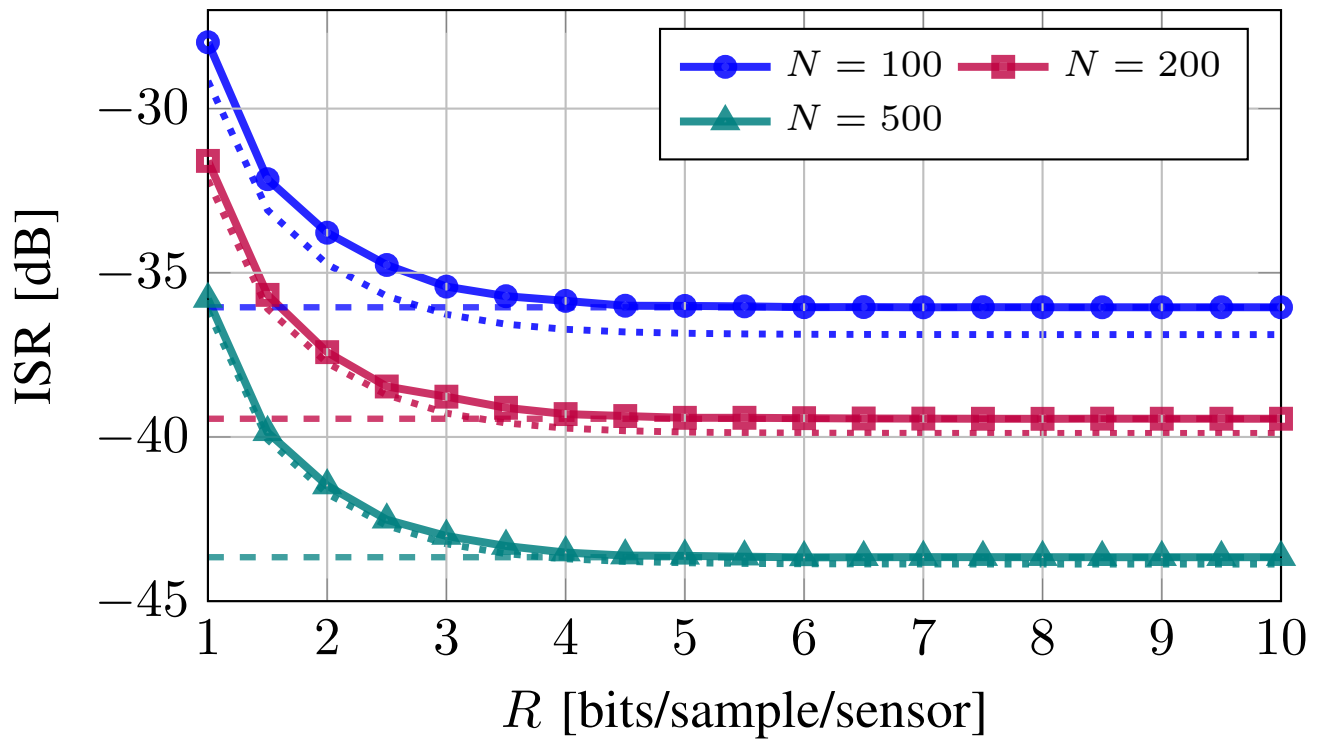
\begin{figure}[!t]
    \centering
    \begin{tikzpicture}
        \begin{axis}[
            width=0.98\linewidth,
            height=0.62\linewidth,
            grid=major,
            xlabel={$R$ [bits/sample/sensor]},
            ylabel={ISR [dB]},
            xmin=1, xmax=10,
            ymin=-45, ymax=-27,
            xtick={1,2,3,4,5,6,7,8,9,10},
            legend pos=north east,
            legend cell align={left},
            legend style={font=\scriptsize},
            legend columns=2,
            axis on top,
        ]

        \addplot [color=blue, mark=o, mark size=1.5pt, line width=1.5pt, opacity=0.85]
            table [x index=0, y index=1] {data/exp1_ml_isr_n100.dat};
        \addplot [color=purple, mark=square, mark size=1.5pt, line width=1.5pt, opacity=0.80]
            table [x index=0, y index=1] {data/exp1_ml_isr_n200.dat};
        \addplot [color=teal, mark=triangle, mark size=1.8pt, line width=1.5pt, opacity=0.85]
            table [x index=0, y index=1] {data/exp1_ml_isr_n500.dat};

        \addplot [dashed, color=blue, line width=1.2pt, opacity=0.75]
            table [x index=0, y index=1] {data/exp1_ml_isr_baseline_n100.dat};
        \addplot [dashed, color=purple, line width=1.2pt, opacity=0.75]
            table [x index=0, y index=1] {data/exp1_ml_isr_baseline_n200.dat};
        \addplot [dashed, color=teal, line width=1.2pt, opacity=0.75]
            table [x index=0, y index=1] {data/exp1_ml_isr_baseline_n500.dat};

        \addplot [dotted, color=blue, line width=1.3pt, opacity=0.85]
            table [x index=0, y index=1] {data/exp1_ml_crlb_n100.dat};
        \addplot [dotted, color=purple, line width=1.3pt, opacity=0.80]
            table [x index=0, y index=1] {data/exp1_ml_crlb_n200.dat};
        \addplot [dotted, color=teal, line width=1.3pt, opacity=0.85]
            table [x index=0, y index=1] {data/exp1_ml_crlb_n500.dat};

        \addlegendimage{no markers, solid, color=blue, line width=1.5pt}
        \addlegendimage{no markers, solid, color=purple, line width=1.5pt}
        \addlegendimage{no markers, solid, color=teal, line width=1.5pt}
        \legend{$N=100$, $N=200$, $N=500$}
        \end{axis}
    \end{tikzpicture}
    \caption{Experiment~1: average ISR versus rate for the Gaussian WSS benchmark. For each color (fixed $N$), solid lines with markers denote the RD-compressed{ oracle-initialized} ML estimator, dashed lines denote the corresponding uncompressed ML baseline, and dotted lines denote the iCRLB on the ISR.}
    \label{fig:exp1_isr_vs_rate}\vspace{-0.5cm}
\end{figure}

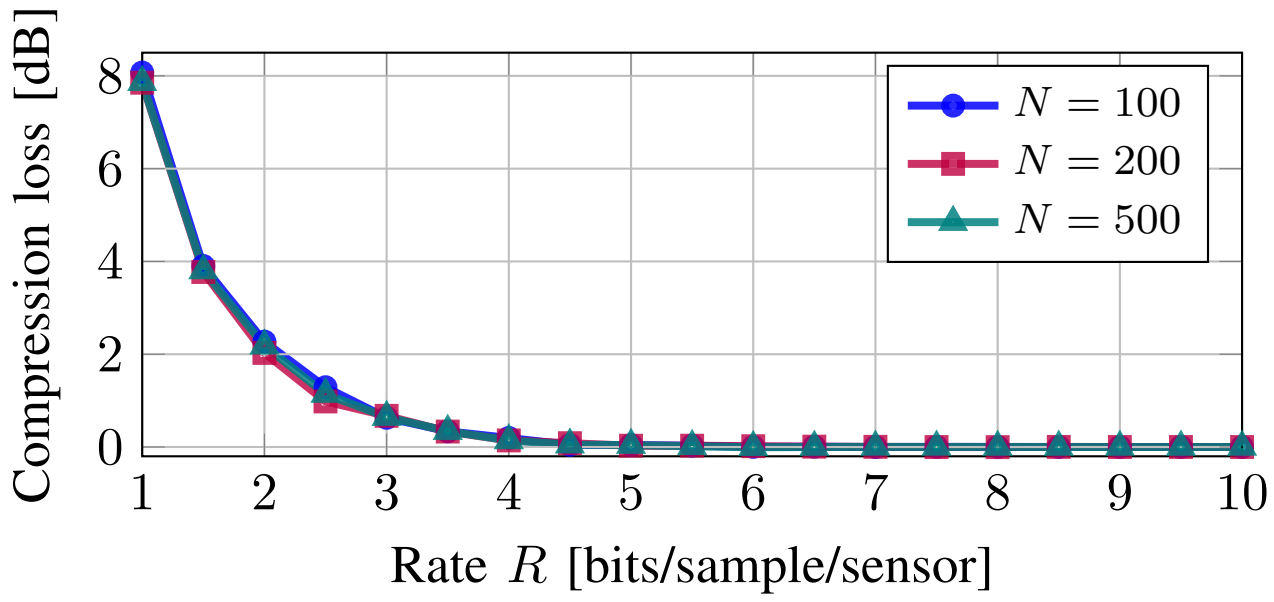
\begin{figure}[t]
    \centering
    \begin{tikzpicture}
        \begin{axis}[
            width=\linewidth,
            height=0.48\linewidth,
            grid=major,
            xlabel={Rate $R$ [bits/sample/sensor]},
            ylabel={Compression loss [dB]},
            xmin=1, xmax=10,
            ymin=-0.2, ymax=8.5,
            xtick={1,2,3,4,5,6,7,8,9,10},
            legend pos=north east,
            legend cell align={left},
            legend style={font=\footnotesize},
            axis on top,
        ]

        \addplot [color=blue, mark=o, mark size=1.5pt, line width=1.5pt, opacity=0.85]
            table [x index=0, y index=1] {data/exp1_ml_gap_db_n100.dat};
        \addplot [color=purple, mark=square, mark size=1.5pt, line width=1.5pt, opacity=0.80]
            table [x index=0, y index=1] {data/exp1_ml_gap_db_n200.dat};
        \addplot [color=teal, mark=triangle, mark size=1.8pt, line width=1.5pt, opacity=0.85]
            table [x index=0, y index=1] {data/exp1_ml_gap_db_n500.dat};

        \legend{$N=100$, $N=200$, $N=500$}
        \end{axis}
    \end{tikzpicture}\vspace{-0.2cm}
    \caption{Experiment~1: compression loss versus rate, computed as the ISR difference in dB between the RD-compressed ML estimator and the uncompressed ML baseline. This view isolates the penalty due solely to rate-limited compression, and reflects the weak dependence on $N$~\eqref{eq:isr_ratio_asymptotic}.}\vspace{-0.2cm}
    \label{fig:exp1_gap_vs_rate}
\end{figure}

Figures~\ref{fig:exp1_isr_vs_rate}--\ref{fig:exp1_gap_vs_rate} show that the
compressed ML performance improves smoothly with $R$, tracks the CRLB, and
the ML--CRLB gap decreases as the sample size increases. The compression loss
is about $8$~dB at $R=1$, falls below $0.2$~dB by $R=4$, and becomes
negligible at high rates. As expected, increasing $N$ shifts all curves
downward in this matched Gaussian setting. Moreover,
Fig.~\ref{fig:exp1_gap_vs_rate} is consistent with
\eqref{eq:isr_ratio_asymptotic}: RD compression changes the asymptotic
constant but not the $1/N$ decay rate, so compression loss depends only weakly
on $N$.

\vspace{-0.3cm}
\subsection{Practical RF-Like Sources Under Dithered Quantization}

The second experiment focuses on the practical compression front-end of Section~\ref{subsec:compression_frontend}. We again take $K=L=3$, but now use a different full-rank, well-conditioned mixing matrix,
\begin{equation}
\matA=\begin{bmatrix}
0.73 & 0.49 & -0.56\\
-0.61 & 0.80 & 0.26\\
0.31 & -0.45 & 0.84
\end{bmatrix}.
\end{equation}
This matrix is deliberately not close to the identity, while remaining far from pathological ill-conditioning. The sources are  non-Gaussian, and protocol-inspired. Specifically, the three source processes are BLE-like GFSK, IEEE~802.15.4-like OQPSK, and WiFi-like OFDM waveforms, generated according to Table~\ref{tab:exp2_sources}~\cite{bluetooth_core_v54,ieee_802154_2020,ieee_80211_2020}.\footnote{In Table~\ref{tab:exp2_sources}, $\mathrm{sps}$ stands for samples per symbol and $BT$ denotes the bandwidth--time product of the Gaussian pulse-shaping filter, where $B$ is the 3-dB bandwidth and $T$ is the symbol duration.} Each waveform is driven by a random data stream, then randomized by a uniformly distributed timing offset over its natural period and by a uniform random carrier phase so that the observed source block is modeled as WSS with a well-defined PSD. The blocklength is $N=4096$, the observation SNR is $20$~dB, and we average over $1000$ MC trials for each rate $R\in\{1,\ldots,10\}$.

\begin{table}[t]
    \centering
    \caption{Experiment~2 source models. Normalized carrier frequencies are in cycles/sample.}
    \label{tab:exp2_sources}
    \footnotesize
    \setlength{\tabcolsep}{3.2pt}
    \renewcommand{\arraystretch}{1.03}
    \begin{tabular}{@{}c l p{0.53\linewidth}@{}}
        \hline
        Source & Model & Parameters \\
        \hline
        1 & BLE-like GFSK & $\mathrm{sps}=8$, $BT=0.5$, Gaussian span $4$ symbols, modulation index $0.5$, carrier $0.11$. \\
        2 & 802.15.4-like OQPSK & $\mathrm{sps}=8$, half-sine pulse shaping, carrier $0.18$. \\
        3 & WiFi-like OFDM & $N_{\mathrm{FFT}}=16$, cyclic prefix $4$, active subcarriers $\pm1,\pm2,\pm3$, QPSK, carrier $0.20$. \\
        \hline
    \end{tabular}\vspace{-0.4cm}
\end{table}

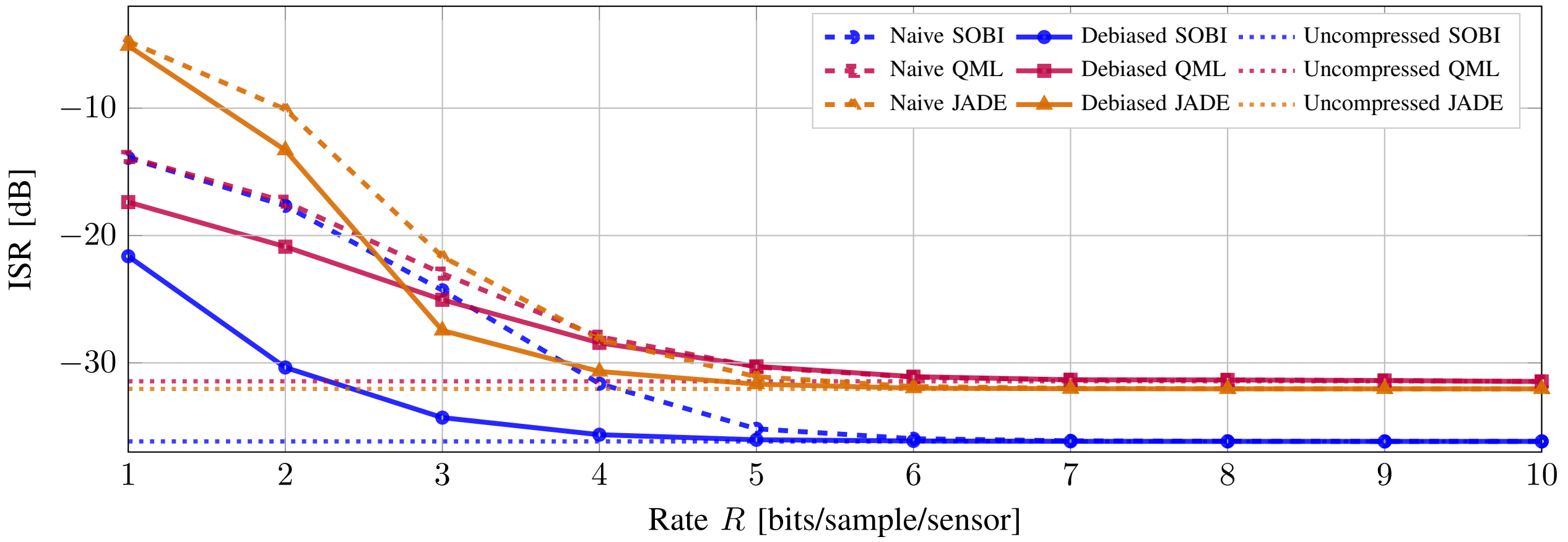
\begin{figure*}[!t]
    \centering
    \begin{tikzpicture}
        \begin{axis}[
            width=0.92\textwidth,
            height=0.35\textwidth,
            grid=major,
            xlabel={Rate $R$ [bits/sample/sensor]},
            ylabel={ISR [dB]},
            xmin=1, xmax=10,
            ymin=-37, ymax=-2,
            xtick={1,2,3,4,5,6,7,8,9,10},
            legend style={font=\scriptsize, at={(0.985,0.985)}, anchor=north east, fill=white, draw=gray!40},
            legend cell align={left},
            legend columns=3,
            axis on top,
        ]

        \addplot [dashed, color=blue, mark=o, mark size=1.4pt, line width=1.5pt, opacity=0.85]
            table [x index=0, y index=1] {data/exp2_sobi_naive_isr.dat};
        \addplot [color=blue, mark=o, mark size=1.4pt, line width=1.5pt, opacity=0.85]
            table [x index=0, y index=1] {data/exp2_sobi_debiased_isr.dat};
        \addplot [dotted, color=blue, line width=1.3pt, opacity=0.75]
            table [x index=0, y index=1] {data/exp2_sobi_uncomp_isr.dat};

        \addplot [dashed, color=purple, mark=square, mark size=1.4pt, line width=1.5pt, opacity=0.82]
            table [x index=0, y index=1] {data/exp2_qml_naive_isr.dat};
        \addplot [color=purple, mark=square, mark size=1.4pt, line width=1.5pt, opacity=0.82]
            table [x index=0, y index=1] {data/exp2_qml_debiased_isr.dat};
        \addplot [dotted, color=purple, line width=1.3pt, opacity=0.75]
            table [x index=0, y index=1] {data/exp2_qml_uncomp_isr.dat};

        \addplot [dashed, color=orange!85!black, mark=triangle, mark size=1.6pt, line width=1.5pt, opacity=0.90]
            table [x index=0, y index=1] {data/exp2_jade_naive_isr.dat};
        \addplot [color=orange!85!black, mark=triangle, mark size=1.6pt, line width=1.5pt, opacity=0.90]
            table [x index=0, y index=1] {data/exp2_jade_debiased_isr.dat};
        \addplot [dotted, color=orange!85!black, line width=1.3pt, opacity=0.75]
            table [x index=0, y index=1] {data/exp2_jade_uncomp_isr.dat};

        \legend{Naive SOBI, Debiased SOBI, Uncompressed SOBI, Naive QML, Debiased QML, Uncompressed QML, Naive JADE, Debiased JADE, Uncompressed JADE}
        \end{axis}
    \end{tikzpicture}
    \caption{Experiment~2: average ISR versus rate for the protocol-inspired RF setting under subtractively dithered quantization. The plot compares SOBI, the practical alternating Gaussian-likelihood benchmark labeled QML, and JADE, each in naive, debiased, and uncompressed forms.}
    \label{fig:exp2_isr_vs_rate}\vspace{-0.2cm}
\end{figure*}

Compression is performed independently at each sensor using subtractively dithered uniform scalar quantization. The per-sensor dynamic range is selected as $\mybar{x}_\ell=\widehat\sigma_\ell\sqrt{2\log(KN^2)}$ from the observed block,\footnote{Here $\widehat{\sigma}_\ell$ denotes the sample standard deviation of the observed block at sensor $\ell$, namely
$
\widehat{\sigma}_\ell
=
\sqrt{
\frac{1}{N}\sum_{n=1}^N
\bigl(\rndx_\ell[n]-\mybar{\rndx}_\ell\bigr)^2
},
\mybar{\rndx}_\ell=\frac{1}{N}\sum_{n=1}^N \rndx_\ell[n].
$} and the mean overload probability in the reported operating range remains negligible (about $1.68\times 10^{-5}$ at $R=1$ and numerically zero for $R\geq 2$). At the FC we compare three separation families on the same compressed observations: SOBI with $20$ lags, JADE, and a Gaussian QML benchmark. The QML routine assumes a Gaussian frequency-domain covariance model
\begin{equation}
\matC_k=\matA \matP_k \tp{\matA}+\matSigma_{\mathrm{noise}},
\end{equation}
where $\matP_k$ is the diagonal matrix of source PSDs at frequency bin $k$, and $\matSigma_{\mathrm{noise}}$ denotes the sensor-domain covariance of the additive noise term. For the uncompressed observations, $\matSigma_{\mathrm{noise}}=\matR_{\rvecw}$, whereas for the dithered compressed observations, $\matSigma_{\mathrm{noise}}=\matR_{\rvecw} + \matR_{\rvece}$, with $\matR_{\rvece}=\diag\left(\Delta_1^2/12,\ldots,\Delta_K^2/12\right)$. The QML is initialized from the SOBI estimate, then alternates between estimating the source PSDs from the current demixed signals, and updating $\matA$ by Fisher-scoring steps on the resulting Gaussian quasi-log-likelihood. In this implementation, we use $5$ outer alternations and $16$ inner Fisher-scoring steps per alternation. For each family, we compare the naive original separation rule to its compression-aware debiased counterpart, together with the corresponding uncompressed baseline. {For numerical stability, the debiased covariance is symmetrized and its eigenvalues are floored at $10^{-8}$ to ensure positive definiteness.}

\begin{figure}[t]
    \begin{tikzpicture}
        \begin{axis}[
            width=\linewidth,
            height=0.5\linewidth,
            grid=major,
            xlabel={Rate $R$ [bits/sample/sensor]},
            ylabel={Debiasing gain [dB]},
            xmin=1, xmax=10,
            ymin=0, ymax=14.5,
            xtick={1,2,3,4,5,6,7,8,9,10},
            legend pos=north east,
            legend cell align={left},
            legend style={font=\footnotesize},
            axis on top,
        ]

        \addplot [color=blue, mark=o, mark size=1.4pt, line width=1.5pt, opacity=0.85]
            table [x index=0, y index=1] {data/exp2_gain_sobi_debiasing.dat};
        \addplot [color=purple, mark=square, mark size=1.4pt, line width=1.5pt, opacity=0.82]
            table [x index=0, y index=1] {data/exp2_gain_qml_debiasing.dat};
        \addplot [color=orange!85!black, mark=triangle, mark size=1.6pt, line width=1.5pt, opacity=0.90]
            table [x index=0, y index=1] {data/exp2_gain_jade_debiasing.dat};

        \legend{SOBI, QML, JADE}
        \end{axis}
    \end{tikzpicture}\vspace{-0.15cm}
    \caption{Experiment~2: debiasing gain, defined as the ISR difference in dB between the naive and debiased versions of each separation family. Positive values indicate that the compression-aware correction improves separation. SOBI exhibits the largest low-rate gain, QML shows a smaller but consistent gain, and JADE reaches its strongest gain at intermediate low rates.}
    \label{fig:exp2_debiasing_gain}\vspace{-0.35cm}
\end{figure}
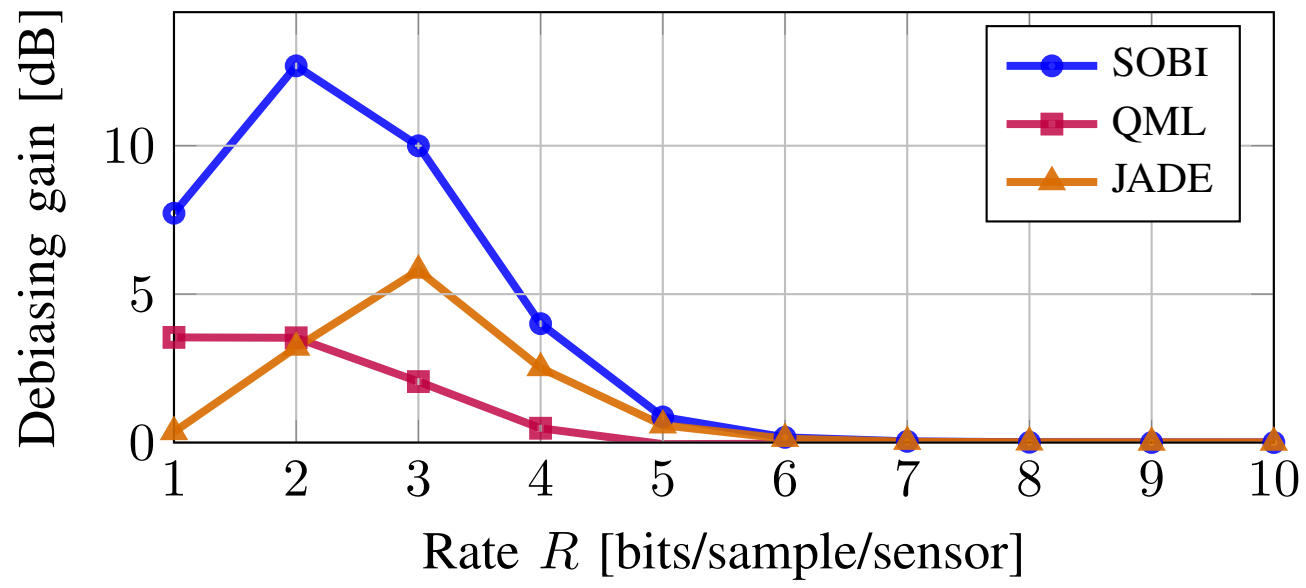

Figure~\ref{fig:exp2_isr_vs_rate} shows that dithered-quantization-compensating debiasing is indeed essential. For SOBI, debiasing improves the average ISR from about $-13.9$~dB to $-21.6$~dB at $R=1$ and from about $-17.7$~dB to $-30.4$~dB at $R=2$, before approaching the uncompressed reference around $-36.2$~dB. QML exhibits the same qualitative trend, although with smaller gains and weaker overall performance than SOBI in this implementation. JADE benefits, generally less, mainly at intermediate low rates: its debiasing gain is about $3.2$~dB at $R=2$, peaks near $5.8$~dB at $R=3$, and remains above $2.5$~dB at $R=4$. Figure~\ref{fig:exp2_debiasing_gain} summarizes these effects directly.

\vspace{-0.35cm}
\subsection{Distributed Industrial Acoustic Monitoring}
\begin{figure*}[t]
    \centering
    \subfloat[Gaussian phase map\label{fig:exp3_gaussian_phase_heatmap}]{%
    \begin{minipage}[t]{0.495\textwidth}
        \centering
        \begin{tikzpicture}
        \begin{axis}[
            width=0.88\linewidth,
            height=0.72\linewidth,
            xlabel={Rate $R$ [bits/sample/sensor]},
            ylabel={Difficulty parameter $\epsilon$},
            xmin=0.5, xmax=5.0,
            ymin=0.001, ymax=0.010,
            xtick={0.5,1,2,3,4,5},
            ytick={0.001,0.003,0.005,0.007,0.009},
            grid=major,
            grid style={gray!20},
            colorbar,
            point meta min=0,
            point meta max=1,
            colorbar style={ylabel={Success probability}, ytick={0,0.2,0.4,0.6,0.8,1.0}, font=\footnotesize},
            colormap/viridis,
            axis on top,
        ]
        \addplot [matrix plot*, mesh/rows=19, mesh/cols=10, point meta=explicit, draw=none, opacity=0.88] table [x index=0, y index=1, meta index=2] {data/exp3a_heatmap_success.dat};
        \addplot [white, line width=1.2pt, unbounded coords=jump] table [x index=0, y index=1] {data/exp3a_success_boundary_50.dat};
        \node[font=\small\bfseries, text=black] at (axis cs:1.25,0.00225) {$-10.2$};
        \node[font=\small\bfseries, text=black] at (axis cs:2.25,0.00475) {$-11.2$};
        \node[font=\small\bfseries, text=black] at (axis cs:3.25,0.00675) {$-12.5$};
        \node[font=\small\bfseries, text=black] at (axis cs:4.25,0.00875) {$-14.4$};
        \end{axis}
        \end{tikzpicture}
    \end{minipage}}
    \hfill
    \subfloat[Representative slices\label{fig:exp3_gaussian_phase_slices}]{%
    \begin{minipage}[t]{0.495\textwidth}
        \centering
        \begin{tikzpicture}
        \begin{axis}[
            width=0.88\linewidth,
            height=0.72\linewidth,
            xlabel={Rate $R$ [bits/sample/sensor]},
            ylabel={Success probability},
            xmin=0.5, xmax=5.0,
            ymin=-0.02, ymax=1.02,
            xtick={0.5,1,2,3,4,5},
            ytick={0,0.2,0.4,0.6,0.8,1.0},
            grid=major,
            grid style={gray!20},
            legend pos=south east,
            legend cell align={left},
            legend style={font=\scriptsize},
            legend columns=2,
            axis on top,
        ]
        \addplot [color=red!75!black, mark=o, mark size=1.4pt, line width=1.4pt] table [x index=0, y index=1] {data/exp3a_success_slices.dat};
        \addplot [color=blue!80!black, mark=square, mark size=1.4pt, line width=1.4pt] table [x index=0, y index=2] {data/exp3a_success_slices.dat};
        \addplot [color=green!55!black, mark=triangle, mark size=1.6pt, line width=1.4pt] table [x index=0, y index=3] {data/exp3a_success_slices.dat};
        \addplot [color=orange!85!black, mark=diamond, mark size=1.5pt, line width=1.4pt] table [x index=0, y index=4] {data/exp3a_success_slices.dat};
        \legend{$\epsilon=0.001$,$\epsilon=0.0045$,$\epsilon=0.0065$,$\epsilon=0.01$}
        \end{axis}
        \end{tikzpicture}
    \end{minipage}}
    \caption{Experiment~3A: phase transition under the Gaussian WSS surrogate for distributed industrial acoustic monitoring. (a) success-probability heat map over rate and spectral spacing $\epsilon$, with the white contour indicating the empirical $50\%$ mid-transition curve. Black annotations report representative average ISR values in dB at selected grid points. (b) representative success-probability $\epsilon$-slices.}
    \label{fig:exp3_gaussian_phase}\vspace{-0.35cm}
\end{figure*}
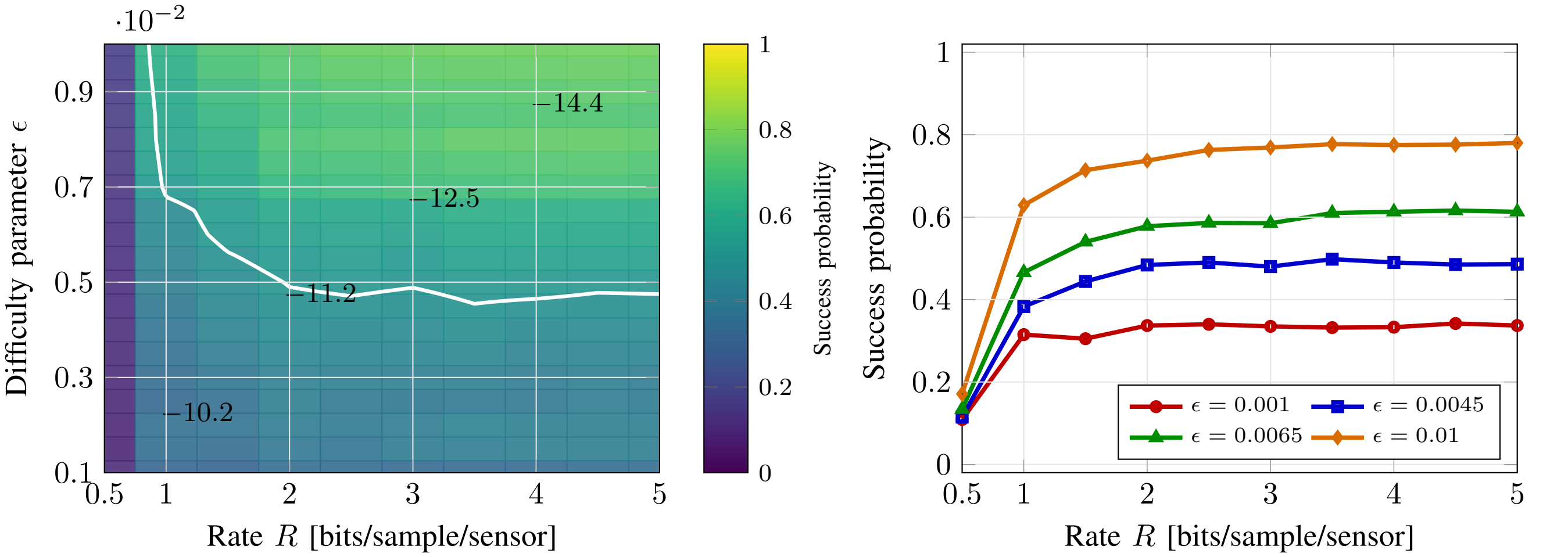

The third experiment is built around a wireless acoustic sensor network for distributed industrial condition monitoring. Multiple spatially separated sensors observe linear mixtures of several simultaneously operating machines, compress their local measurements, and forward only rate-limited descriptions to a FC. We use this setting to illustrate an operational transition between unreliable and usable separation.

We begin with a controlled Gaussian WSS surrogate under the RD compression model. The distributed mixtures are generated with the mixing matrix
\begin{equation}\label{eq:mixingmatrix3}
\matA=\begin{bmatrix}
0.91 & -0.63 & 0.27\\
0.36 & 0.69 & -0.58\\
-0.42 & 0.51 & 0.77
\end{bmatrix}.
\end{equation}
Two sources are narrowband machine-signature surrogates with center frequencies at $f_0\pm\epsilon$, while the third source is a spectrally distinct colored interferer. More precisely, the three one-sided source PSDs are normalized Gaussian bumps: the first two are centered at $0.18\pm\epsilon$ with width $0.035$, and the third is centered at $0.34$ with width $0.06$. We fix $K=L=3$, $N=100$,{\footnote{{{The short blocklength $N=100$ is used to examine a finite-sample
regime in which the joint effect of rate and spectral separability is visible within the simulated grid.}}}} and an observation SNR of $10$~dB, sweep the rate over $R\in\{0.5,1,1.5,\ldots,5\}$ bits/sample/sensor, sweep $\epsilon$ over $\{0.001,0.0015,\ldots,0.01\}$, and average over $1000$ independent MC trials per grid point. At each grid point, the FC applies the same local Gaussian ML estimator (via Fisher scoring) used in Experiment~1, matched to the corresponding compressed Gaussian covariance model. A trial is declared successful when the average ISR falls below $-15$~dB. Figure~\ref{fig:exp3_gaussian_phase_heatmap} displays the resulting success-probability map. The white contour is the \emph{mid-transition} curve, namely the set of points at which the empirical success probability reaches $50\%$, which serves as an (arbitrarily chosen) reliable-operation boundary. The contour, obtained by interpolating the gridded empirical success probabilities, enters the plotted region near $\epsilon\approx 6.8\times 10^{-3}$ at $R=1$, drops to about $\epsilon\approx 4.9\times 10^{-3}$ by $R=2$, and then remains in a relatively narrow band around $(4.7$--$4.9)\times 10^{-3}$ for $R\geq 2.5$. Figure~\ref{fig:exp3_gaussian_phase_slices} shows representative slices at $\epsilon\in\{0.001,0.0045,0.0065,0.01\}$: a phase transition occurs at low rates, where success rises most rapidly, and then continues to rise more gradually.

We then replace the Gaussian surrogates by real industrial machine sounds from the public malfunctioning industrial machine investigation and inspection (MIMII) dataset~\cite{purohit2019mimii}, while keeping the same distributed sensing architecture and the same mixing matrix \eqref{eq:mixingmatrix3}. For the MIMII-based real-data experiment, we first consider only recordings labeled as normal operating conditions. For each machine instance in the dataset, identified by its machine type and ID (e.g., \texttt{fan:id\_04}), we form a normalized log-PSD signature from a representative recording. We then select, as a challenging source pair, the two instances of the same machine type with the most similar signatures, and choose as the third source an instance from a different machine type with a spectrally distinct signature. In our simulation, this procedure selected \texttt{fan:id\_04}, \texttt{fan:id\_02}, and \texttt{pump:id\_04}; the corresponding recordings were used as the three latent sources.

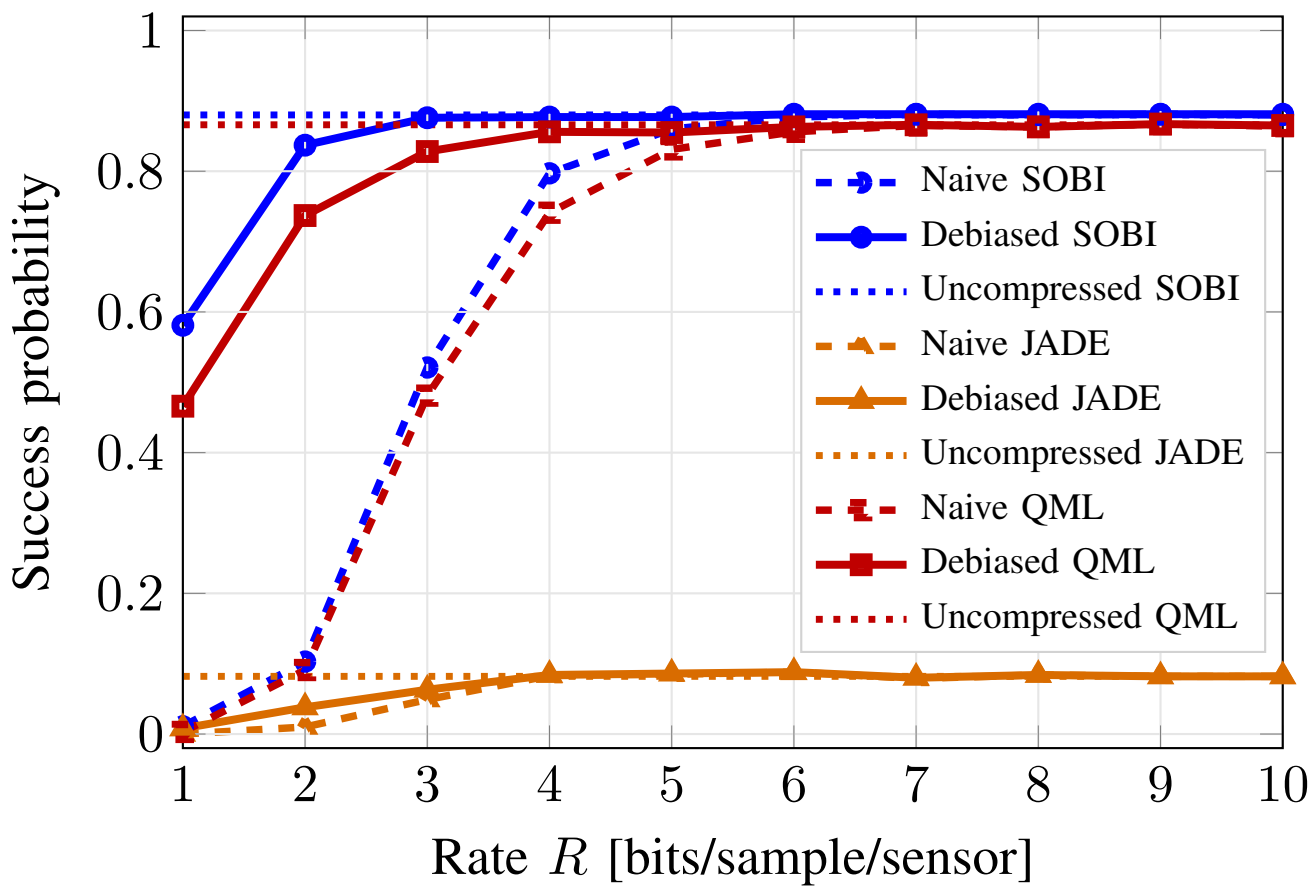
\begin{figure}[t]
    \centering
    \begin{tikzpicture}
    \begin{axis}[
        width=\linewidth,
        height=0.725\linewidth,
        xlabel={Rate $R$ [bits/sample/sensor]},
        ylabel={Success probability},
        xmin=1, xmax=10,
        ymin=-0.02, ymax=1.02,
        xtick={1,2,3,4,5,6,7,8,9,10},
        ytick={0,0.2,0.4,0.6,0.8,1.0},
        grid=major,
        grid style={gray!20},
        legend cell align={left},
        legend style={font=\scriptsize, at={(0.985,0.13)}, anchor=south east, fill=white, draw=gray!35, inner xsep=2pt, inner ysep=1pt},
        legend columns=1,
        axis on top,
    ]
    \addplot [color=blue, mark=o, mark size=1.4pt, line width=1.4pt, dashed] table [x index=0, y index=1] {data/exp3b_sobi_success_naive.dat};
    \addplot [color=blue, mark=o, mark size=1.4pt, line width=1.5pt] table [x index=0, y index=1] {data/exp3b_sobi_success_debiased.dat};
    \addplot [color=blue, line width=1.3pt, dotted] table [x index=0, y index=1] {data/exp3b_sobi_success_unc.dat};
    \addplot [color=orange!85!black, mark=triangle, mark size=1.5pt, line width=1.4pt, dashed] table [x index=0, y index=1] {data/exp3b_jade_success_naive.dat};
    \addplot [color=orange!85!black, mark=triangle, mark size=1.5pt, line width=1.5pt] table [x index=0, y index=1] {data/exp3b_jade_success_debiased.dat};
    \addplot [color=orange!85!black, line width=1.3pt, dotted] table [x index=0, y index=1] {data/exp3b_jade_success_unc.dat};
    \addplot [color=red!75!black, mark=square, mark size=1.4pt, line width=1.4pt, dashed] table [x index=0, y index=1] {data/exp3b_qml_success_naive.dat};
    \addplot [color=red!75!black, mark=square, mark size=1.4pt, line width=1.5pt] table [x index=0, y index=1] {data/exp3b_qml_success_debiased.dat};
    \addplot [color=red!75!black, line width=1.3pt, dotted] table [x index=0, y index=1] {data/exp3b_qml_success_unc.dat};
    \legend{Naive SOBI, Debiased SOBI, Uncompressed SOBI, Naive JADE, Debiased JADE, Uncompressed JADE, Naive QML, Debiased QML, Uncompressed QML}
    \end{axis}
    \end{tikzpicture}\vspace{-0.3cm}
    \caption{Experiment~3B: corroboration on a hard triplet of industrial machine sounds from the MIMII dataset. The same low-rate transition seen in the Gaussian surrogate remains visible for both SOBI and QML on real machine sounds, while JADE exhibits only a modest benefit on this particular triplet. {The source waveforms are real recordings, whereas the three sensor channels are synthetic instantaneous mixtures.}}\vspace{-0.4cm}
    \label{fig:exp3_realistic_mimii}
\end{figure}

In each trial, we draw one random file and one random segment for each selected machine instance. The resulting segments are mean-removed and variance-normalized, and are used as the three latent source signals before synthetic instantaneous mixtures are generated across the three sensors. Each sensor is quantized by subtractively dithered scalar quantization at integer rates, and the FC applies SOBI with $20$ lags, JADE, and the alternating Gaussian-QML, each with and without the relevant debiasing correction. We use $N=4096$, $1000$ MC trials per rate, and again declare success when the average ISR falls below $-15$~dB. Figure~\ref{fig:exp3_realistic_mimii} shows that, on this triplet, debiasing materially shifts the low-rate transition for both SOBI and QML: for SOBI the success probability improves from $1.1\%$ to $58.1\%$ at $R=1$ and from $10.3\%$ to $83.7\%$ at $R=2$, while for QML it improves from $0.3\%$ to $46.6\%$ at $R=1$ and from $9.1\%$ to $73.7\%$ at $R=2$, with the debiased curve nearly matching the uncompressed QML reference by $R=4$. JADE remains weak on this triplet, with only modest gains from debiasing and an uncompressed success probability of only $8.2\%$. Thus, the discriminative structure in this example is predominantly SOS. The mean overload rate remains negligible, approximately $1.50\times 10^{-4}$ at $R=1$ and $2.4\times 10^{-6}$ at $R=2$, so the observed transition is driven by information loss rather than clipping artifacts.

\vspace{-0.4cm}
\section{Concluding Remarks}\label{sec:conclusion}

This paper studied BSS when the mixture observations are not available at the FC in raw form, but only through rate-limited compressed descriptions generated separately at multiple sensing nodes. This perspective brings classical BSS into direct contact with modern distributed sensing architectures, in which communication, storage, and energy constraints are often as fundamental as estimation accuracy itself. Within this setting, we developed a general formulation for compression-aware separation, established identifiability viewpoints under compressed observations, introduced oracle-normalized performance benchmarks, derived ISR lower bounds for the Gaussian WSS case under RD-optimal compression, and proposed practical debiasing corrections for both fourth-order cumulant-based and SOS-based methods. Simulation results show that compression-aware processing recovers a substantial fraction of the separation performance otherwise lost to compression.

This work makes clear that distributed BSS under joint compression and separation raises questions that do not arise in the classical centralized setting. In a centralized architecture, one typically asks what can be inferred from the full sensor data once they have been collected. In the distributed setting considered here, one must also ask what information about separability survives local compression, how encoder decisions affect identifiability, and how a limited communication budget should be allocated across sensing nodes and across time. This suggests a broader research agenda in which BSS is treated as an inverse
problem under communication constraints.

Several theoretical directions appear especially important. First, beyond the Gaussian RD benchmark studied here, it is natural to seek sharper converse and achievability characterizations for separation under finite per-sensor rate budgets, including task-aware bounds stated directly in terms of ISR or related separation metrics. Second, nonasymptotic analyses are needed: finite-block effects, model mismatch, nuisance parameters, and partial prior knowledge can all affect the compression--separation tradeoff, especially in low-rate regimes. Third, identifiability under compression should be understood more broadly for non-Gaussian, nonstationary, or convolutive mixture models that arise in wireless sensing, industrial monitoring, and immersive audio. Finally, it would be valuable to connect information-theoretic limits with computational limits, clarifying when a favorable compression level is sufficient only in principle and when it is accompanied by practical algorithms that can attain the predicted performance.

{The present formulation is restricted to instantaneous linear mixtures. In particular, the MIMII experiment uses real single-channel machine recordings as source waveforms but generates the sensor observations through a synthetic instantaneous matrix. Extending the model and the compression-aware corrections to propagation delays, room impulse responses, and convolutive mixtures is therefore an important next step.}

On the algorithmic side, many challenges remain open as well. In this paper, separation is carried out centrally after the compressed observations are delivered to the FC. A natural next step is to study genuinely distributed separation schemes, in which part of the demixing computation itself is performed across the network through local message passing, iterative consensus, federated updates, or hierarchical aggregation. Such schemes would better reflect large-scale sensor systems in which transmitting all compressed samples to one node may itself be impractical. Another important direction is online and adaptive operation, where both the source statistics and the communication budget may vary over time.

Taken together, these challenges and opportunities suggest that compression-aware distributed BSS should be viewed as an important and largely open extension of classical BSS, with its own identifiability questions, performance limits, and algorithmic opportunities. The framework developed here may provide a starting point for this line of work, and may motivate a broader view of BSS in which communication, computation, and separation are designed jointly rather than in isolation.

\vspace{-0.25cm}
\bibliographystyle{IEEEbib}
\bibliography{./Inputs/refs}

\end{document}